\documentclass{elsarticle}
\usepackage[utf8]{inputenc}
\usepackage[margin=1in]{geometry}
\usepackage[font=normalsize]{caption}
\usepackage{parskip}
\usepackage{subfigure}
\usepackage{hyperref}
\usepackage{xcolor}
\usepackage{xspace}
\usepackage{aas_macros}
\hypersetup{urlcolor=blue, colorlinks=true,citecolor=blue}
\graphicspath{{figures/}}

\definecolor{purp}{RGB}{140,0,255}

\newcommand{\ml}[1]{{\textcolor{black}{#1}}}
\newcommand{\astronomaly}{\textsc{Astronomaly}\xspace}
\newcommand{\gap}{\vspace{15pt}}

\begin{document}
%\title{Astronomaly: A general framework for anomaly detection with active learning in astronomical data}
% \title{Astronomaly: Active Learning for Anomaly Detection in Astronomical Data}
% \title{\astronomaly: A New Approach to Active Anomaly Detection in Astronomical Data}
\title{\astronomaly: Personalised Active Anomaly Detection \\in Astronomical Data}

\author[a,b,c]{Michelle Lochner \corref{correspondingauthor}}
\cortext[correspondingauthor]{Corresponding author}
\ead{mlochner@uwc.ac.za}
\author[b,c,d,e]{Bruce A. Bassett}
\address[a]{Department of Physics and Astronomy, University of the Western Cape, Bellville, Cape Town, 7535, South Africa}
\address[b]{South African Radio Astronomy Observatory (SARAO), The Park, Park Road, Pinelands, Cape Town 7405, South Africa}
\address[c]{African Institute for Mathematical Sciences, 6 Melrose Road, Muizenberg, 7945, South Africa}
\address[d]{Department of Maths and Applied Maths, University of Cape Town, Cape Town, South Africa}
\address[e]{South African Astronomical Observatory, Observatory, Cape Town, 7925, South Africa}

\begin{abstract}
Survey telescopes such as the Vera C. Rubin Observatory and the Square Kilometre Array will discover billions of static and dynamic astronomical sources. Properly mined, these enormous datasets will likely be wellsprings of rare or unknown astrophysical phenomena. The challenge is that the datasets are so large that most data will never be seen by human eyes; currently the most robust instrument we have to detect relevant anomalies. Machine learning is a useful tool for anomaly detection in this regime. However, it struggles to distinguish  between interesting anomalies and irrelevant data such as instrumental artefacts or rare astronomical sources that are simply not of interest to a particular scientist. Active learning combines the flexibility and intuition of the human brain with the raw processing power of machine learning. By strategically choosing specific objects for expert labelling, it minimises the amount of data that scientists have to look through while maximising potential scientific return. Here we introduce \astronomaly: a general anomaly detection framework with a novel active learning approach designed to provide personalised recommendations. \astronomaly can operate on most types of astronomical data, including images, light curves and spectra. We use the Galaxy Zoo dataset to demonstrate the effectiveness of \astronomaly, as well as simulated data to thoroughly test our new active learning approach. We find that for both datasets, \astronomaly roughly doubles the number of interesting anomalies found in the first 100 objects viewed by the user. \astronomaly is easily extendable to include new feature extraction techniques, anomaly detection algorithms and even different active learning approaches. The code is publicly available at \verb!https://github.com/MichelleLochner/astronomaly!. 
\end{abstract}

%\begin{document}

\maketitle

\section{Introduction}
Modern telescopes are taking increasingly large amounts of data resulting in surveys of hundreds to thousands of square degrees containing millions of astronomical objects. Upcoming telescopes such as the Vera C. Rubin Observatory\footnote{\url{https://www.lsst.org/}} \citep{Ivezic2019} and the Square Kilometre Array\footnote{\url{https://www.skatelescope.org/}} will produce reduced datasets on the scale of petabytes. One of the most tantalising prospects of these datasets is the detection of rare and previously undiscovered astrophysical objects and phenomena, known as anomalies. In the past, groundbreaking and unexpected discoveries have been made when a human analysing data noticed something unusual. However as data quantities increase, human monitoring for anomalies becomes intractable.

One solution to this problem is to crowd-source anomaly detection with citizen science. Galaxy Zoo \citep{lintott2008, lintott2011} is an excellent example of citizen science where over 50000 participants have taken part to visually classify thousands of galaxies, including taking note when something odd is observed about the galaxy. While citizen science has a critical role to play in the exploration and exploitation of large datasets, few citizens can replace an expert, particularly when the task is as ill-defined as looking for anomalies that may have never been seen before. \ml{Many astronomical data types, such as spectra or light curves, can be complex and challenging for citizens to interpret.} Additionally, datasets are quickly growing too large for even large teams of citizens to exhaustively study.

Machine learning has become extremely popular in astronomy and other fields as a powerful tool for automating many complex tasks that were previously done by humans. The use of machine learning for anomaly detection has seen a rapid increase, particularly in the past two years, as scientists grapple with expanding data volumes.

Anomaly detection algorithms typically form part of unsupervised learning: a branch of machine learning that operates without labels for the data. There are a diverse range of algorithms that operate on a variety of different principles which attempt to separate outliers from the majority. In astronomy, machine learning algorithms for anomaly detection have been applied to spectra \citep[e.g.][]{skoda2020, Baron_2016}, time series data \citep[e.g.][]{martinezgalarza2020, Giles_2018} and images \citep[e.g.][]{Hocking_2017, doorenbos2020, Margalef_Bentabol_2020}. Particularly worth mentioning is the unsupervised learning package \textsc{PINK} \citep{polsterer2019} that applies self-organising maps to perform clustering and anomaly detection on image data. \ml{As well as unsupervised approaches, many authors have applied semi-supervised machine learning algorithms; where part of the dataset has labels, in the context of anomaly detection usually the ``normal'' dataset is labelled \citep[e.g.][]{Solarz2017,Clarke2020,Marianer2020}.} There have of course been many more successful applications of anomaly detection in astronomical data than have been mentioned here.

While machine learning can be an excellent tool for automated anomaly detection, many of the anomalies detected will not be of interest. This idea of \emph{relevance} is inherently subjective: most astronomers would not be interested in artefacts in data that may be flagged as anomalous, but someone in charge of data quality control would certainly be. Additionally, an anomalous supernova may be of little interest to an astronomer whose focus is on variable stars. Without explicitly designing a classification algorithm to search for a particular type of anomaly (which cannot be done for unknown unknowns), a machine learning algorithm cannot tell the difference between a relevant and irrelevant anomaly, usually resulting in a high false positive rate for the end user.

Active learning  attempts to efficiently combine the intuition and domain knowledge of an expert with the raw processing power of machine learning. The time taken for an expert to examine data is valuable and active learning helps to make optimal use of this time. Active learning has been used in astronomy to greatly reduce the amount of human labelling required for the classification of images \citep{Walmsley_2019}, variable stars \citep{Richards_2011} and a variety of other applications. Active learning has also been applied specifically to anomaly detection for time domain astronomy \citep{ishida2020} by adapting a particular anomaly detection algorithm. \cite{webb2020} also applied active anomaly detection to transient data, making use of an early version of the approach introduced in this paper.

In this work, we introduce a general framework to apply a new personalised active learning approach for anomaly detection in astronomical data. This framework, that we affectionately dub \astronomaly \footnote{\url{https://github.com/MichelleLochner/astronomaly}}, is envisioned to operate on almost any type of astronomical data including images, time series data, spectra and general sets of features. \astronomaly  has a python backend and a JavaScript frontend, to allow for the possibility of being easily accessed on a remote server. The backend includes some data preprocessing, feature extraction and several machine learning anomaly detection algorithms. The frontend allows a user to easily explore the data, applying labels to objects based on how relevant or interesting a particular object is to that user. 

We introduce a novel active learning approach for combining the user-selected relevance score with the machine learning anomaly score to effectively create a recommendation engine, allowing the expert to rapidly focus on particular types of anomalies without needing to know what they are a priori.

Both \cite{reis2019} and \cite{fluke2020} noted the critical importance of the interface used to effectively combine machine learning with human input in unsupervised learning scenarios. We note that \cite{kong2020} independently reached very similar conclusions to us about the need for ``relevance'' labelling, but have an extremely different approach to the problem.

We begin this paper by introducing our novel active learning approach in Section \ref{sec:active_learning} and follow with a high level overview of \astronomaly in Section \ref{sec:framework}. We relegate details of the code to the documentation\footnote{\url{https://astronomaly.readthedocs.io/en/latest/}} and also leave technical details to the Appendix. Section \ref{sec:evaluation} briefly discusses how to evaluate the performance of anomaly detection algorithms. We then demonstrate \astronomaly in Section \ref{sec:galaxy_zoo} on Galaxy Zoo data. Section \ref{sec:simulations} further evaluates our active learning approach using simulations. Finally we conclude in Section \ref{sec:conclusions} and highlight future directions for \astronomaly.

\section{A new approach to active learning}
\label{sec:active_learning}
The goal of our active learning method is to optimise the use of a limited resource: the time spent by a human looking at the data. While machine learning does most of the hard work in detecting anomalies, it cannot {\em a priori} tell the difference between a relevant and an irrelevant anomaly, since this is intrinsically subjective: while one astronomer may be interested in an oddly behaving variable star another may only want to know about new extragalactic transients. If the anomaly is known and has a large corresponding training set, a standard classifier could be used but in the case considered in this work, it is assumed the data are initially completely unlabelled and a user may not have a clear idea of the objects that will be relevant before looking at the data. We therefore use an active learning approach to address this situation.

\ml{Typically active learning (AL) algorithms are discussed in the context of classification problems, where the AL algorithm tries to estimate which examples will lead to the most improvement in the classification performance. In this case we choose to optimise a different cost function. We want to show the human the most interesting objects which then, in turn provide the new leaning data for the algorithm on retraining. This makes our approach closer to a recommendation engine rather than traditional active learning. For a review of more traditional AL methods, see e.g. \cite{settles2009}.}

\subsection{Overview}
As with most machine learning applications, the first step is to extract features, which are typically lower dimensional representations of the data that contain relevant information (see Section \ref{sec:features} and \ref{sec:ellipse}). The next step is to run an anomaly detection algorithm (see Section \ref{sec:anomaly_algorithms}), which allows the data to be sorted according to anomaly score. This already prioritises the more anomalous data for the human expert to go through, but both relevant and irrelevant anomalies will be appear high up in the rankings.

Now the expert must score some of these anomalies by how relevant or interesting each object is. Section \ref{sec:framework} outlines our interactive approach for providing these scores, but the method is general and only requires that relevance scores are provided for a subset of the data. A second layer of machine learning, which is a \emph{regressor} is then run to try to predict relevance scores for new objects not yet viewed by the user. The anomaly detection algorithm only needs to be run once (unless additional data is added) but the relevance score may be recomputed as more data are labelled.

The initial ``training set'' of user-labelled data is likely to be very small, with many parts of feature space that have no labels. Thus we need to find a way to combine the predicted relevance score, the anomaly score \emph{and} the algorithm's confidence in the predicted relevance. In regions where the algorithm is uncertain, it should default back to the machine-learned anomaly score which would mean the user gets the opportunity to see and label the object so new anomalies are less likely to be missed. In regions where confidence is high, the predicted relevance score should dominate meaning that objects similar to what the user has previously labelled as ``irrelevant'' will be ranked lower and not be likely to be seen. We combine these by constructing a \emph{new} trained anomaly score for each object. Our active learning approach therefore consists of two layers of machine learning, one to determine the anomaly score and another to predict the relevance score, combined in an equation that incorporates uncertainty in the relevance score.

\subsection{Combining the anomaly and relevance scores}
To summarise, our aim was to develop a novel active learning method that achieves the following goals:
\begin{itemize}
    \item Able to take user input and adjust anomaly scores according to human-given relevance, to learn which anomalies are interesting. 
    \item Include  uncertainty so that regions of feature space with little nearby human-scored data revert back to the raw anomaly score
    \item Able to work with any anomaly detection and scoring algorithm
\end{itemize}
These goals can be achieved through ranking of objects based on a new score calculated via the equation
\begin{equation}
\label{eq:active_learning}
    \hat{S} = S\, {\rm tanh} \big(\delta - 1 + {\rm arctanh}(\widetilde{U}) \big),
\end{equation}
where $S$ is the raw anomaly score returned from the machine learning algorithm (normalised onto a chosen scale), $\hat{S}$ is the updated active learning anomaly score, $\delta$ is a distance penalty term, that is large for regions where the user score is not well known, and $\widetilde{U}$ is the normalised ``relevance'' score given by user labelling, computed as:
\begin{equation}
\label{eq:userscore}
    \widetilde{U} = \epsilon_1 + \epsilon_2 \left(\frac{U}{U_{\rm max}} \right),
\end{equation}
where $U$ is the input user relevance score, $U_{\rm max}$ is the maximum possible user score which by default is set to  $U_{\rm max} = 5$ for \astronomaly and $\epsilon_1 = 0.1, \epsilon_2 = 0.85$  are normalisation constants. 

This achieves our stated goals above. When either $\delta \gg 1$ or $U$ is close to $U_{max}$, $\tanh \rightarrow 1$ and we revert to the raw anomaly score, either because of lack of nearby human scores ($\delta \gg 1$) or because of high human scores. On the other hand, if $\delta$ is small and $U$ is low, the raw anomaly score is suppressed by the $\tanh$ part of the above equation, which downweights the overall score.  Of course, we could amplify the effect of human scores by adding additional hyperparameters, most noticeably a width parameter for $\tanh$, which we do not include here. \ml{Additionally, there is no special reason why $\tanh$ should be used in place of another function with similar properties, such as a sigmoid. The normalisation constants $\epsilon_1$ and $\epsilon_2$ were chosen to give stable numerics for the full range of possible user scores, since arctanh diverges for arguments of unity. $U$ is chosen to be an integer purely to align with the user interface as humans are most comfortable assigning scores as integers.} 

\subsection{Predicting the relevance score}
There are two key unknowns that we still need to determine in Equations \ref{eq:active_learning} and \ref{eq:userscore}: the user score $U$, which must be defined for all objects in the dataset, and the distance penalty term $\delta$.

$U$ will only be known for a small percentage of the data: the few objects that the user has labelled. These will primarily be objects ranked as highly anomalous according to the machine learning algorithm. To estimate $U$ for the remainder of the dataset, we train a random forest \citep{breiman2001} regression algorithm, with \verb!n_estimators! set to 100. 
% We do not tune the hyperparameters of this regressor, simply setting \verb!n_estimators! to 100. We acknowledge that this a simplistic approach and the regression algorithm can be improved and tuned for any particular problem. However we expect to always be in the regime of very limited data where errors in the estimates are likely to be dominated by uncertainty due to the small training set, not by the choice of hyperparameters. We did experiment with several regression algorithms and found it did not have a large impact on our results.

We quantify the uncertainty of the regression estimate of $U$ by introducing the distance penalty term, $\delta$, above given by:
\begin{equation}
    \label{eq:distance_penalty}
    \delta = {\rm exp}\left(\alpha \frac{d}{d_0}\right).
\end{equation}
In Equation \ref{eq:distance_penalty}, $d$ is the Euclidean distance, in feature space, of the object in question to its nearest human-labelled neighbour. $d_0$ is the mean distance to a labelled neighbour in the dataset and $\alpha$ is a tuning parameter of order one that allows the determination of how much weight should be placed on the underlying anomaly detection algorithm versus the predicted user scores. \ml{An exponential is chosen here to rapidly suppress the score in regions of parameter space with few labels, but other functions could be explored.} For this paper, $\alpha$ has been set to one. $d$ can be computationally expensive to compute and we make use of a KDTree \citep{maneewongvatana1999} implemented in \verb!scipy! \citep{virtanen2020} which is extremely efficient. An alternative would be to estimate the uncertainty in the regression directly, e.g. through the use of Bayesian regression or optimisation. The end result would be similar.  

\subsection{Visualising the active learning anomaly score}
Putting all the pieces together, we can see that if we compute the active learning score for the human-labelled objects, $d$ is zero, $\delta$ is one and $\hat{S}$ will simply be $S\, \widetilde{U}$ - the anomaly score weighted by the normalised user relevance score. Conversely, for objects far from any labelled data, $d$ will be very large, $\delta$ will be similarly large, which means the tanh in Equation \ref{eq:active_learning} will tend towards 1 and $\hat{S}=S$. Regions of feature space that are in between the two extremes will be weighted according to the human relevance score, as regulated by the $\alpha$ tuning parameter.

Figure \ref{fig:sim_features} illustrates the active learning procedure using the data described in Section \ref{sec:simulations}. Here the feature space is visualised in two dimensions as a heat map and it can be seen that by training the algorithm, objects in regions of feature space near irrelevant data are downweighted, allowing more interesting anomalies to be ranked higher. \ml{\ref{sec:astronomaly_flowchart} gives a summary of the entire \astronomaly framework, particularly the interactive active learning process.}

\noindent
\begin{minipage}{1\linewidth}
    \begin{center}
        \begin{minipage}{1\linewidth}
               \includegraphics[width=1\linewidth]{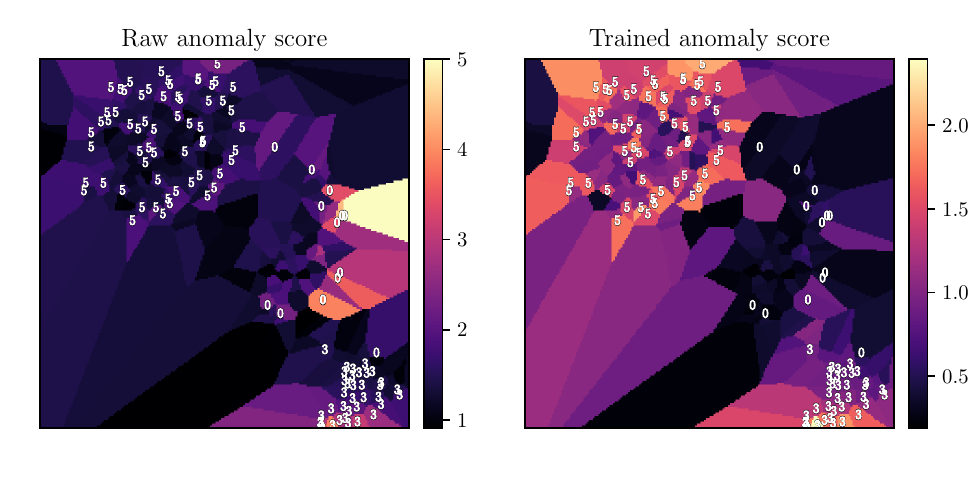}
        \end{minipage}
    \end{center}
    \captionof{figure}{A visualisation of how the active learning process changes the anomaly score in the feature space (a low dimensional representation of the data). The background colour is a t-SNE plot (see Section \ref{sec:visualisation}) that has been converted into a heatmap, coloured by anomaly score before active learning (left) and after (right). The small numbers represent points that have been labelled by a human, with 5 being the most relevant and 0 being the least relevant. Before active learning, the region towards the middle-right of the plot has a high anomaly score but the user labels in that area are zero. This region gets appropriately downweighted after active learning, while the region around the objects scored with a 5 is dramatically upweighted. Objects labelled with a 3, of neutral relevance, do not change significantly. The end result is that objects unseen by the user which are near the 5's in feature space, are now far more likely to be ranked highly and ``recommended'' to the user.}
    \label{fig:sim_features}
\end{minipage}

\section{\astronomaly framework}
\label{sec:framework}
\astronomaly is a general framework for active anomaly detection designed to operate on most kinds of astronomical data. It incorporates machine learning and the active learning approach described in Section \ref{sec:active_learning} and makes use of a python backend and web frontend. Figure \ref{fig:screenshot_anomaly} shows a screenshot of the web interface, written in JavaScript, that allows a user to quickly iterate through their data, sorted by anomaly score, and provide labels of how ``relevant'' a particular object is, thus allowing active learning. \astronomaly's backend consists of a library of data management tools, preprocessing techniques, feature extraction methods and machine learning anomaly detection algorithms (see Figure \ref{fig:flowchart} in \ref{sec:astronomaly_flowchart} for a visual description of the \astronomaly code). 

Each component of \astronomaly can easily be expanded, allowing the introduction of any number of new techniques. Here we briefly overview the capabilities of \astronomaly, leaving the full details (which are liable to change as the code is improved) to the documentation.

\noindent
\begin{minipage}{1\linewidth}
    \centering
    \vspace{20pt}
    \includegraphics[width=1\linewidth]{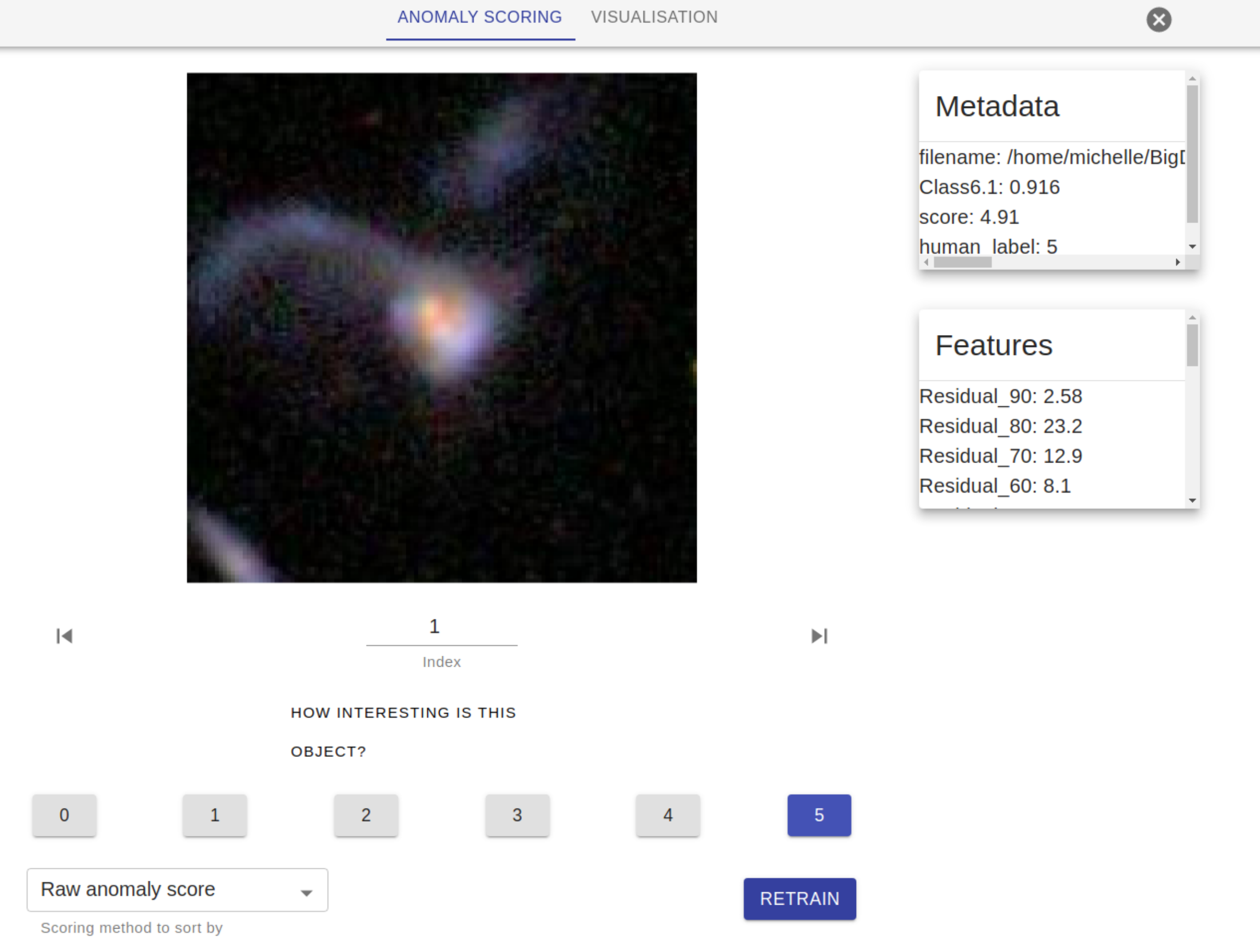}
    \captionof{figure}{A screenshot of the \astronomaly visual interface with a Galaxy Zoo example. This interface is used to quickly scroll through the most anomalous objects labelling them according to relevance.}
    \label{fig:screenshot_anomaly}
\end{minipage} 

\subsection{Data Management}
Our goal is to develop a tool general enough to be able to run on almost any kind of astronomical data. This is only feasible for two reasons: astronomical data is somewhat standardised and, after the feature extraction step (see below), the anomaly detection and active learning algorithms are agnostic to input data type. We have initially focused on image data and one dimensional data such as spectra or unstructured features. Time series data will be fully integrated into \astronomaly in a future version.

In addition to data management tools, \astronomaly features several standard preprocessing algorithms such as log or sinh scaling transforms, sigma clipping and data normalisation. The correct preprocessing of data is often very important for subsequent anomaly detection and requirements can vary between datasets.

\subsection{Feature extraction}
\label{sec:features}
Feature extraction is a critical step of any machine learning pipeline but particularly for anomaly detection. Complex and often high dimensional data can be reduced to simpler sets of features that are thought to incorporate most of the critical information about the data. \astronomaly includes several feature extraction methods such as the Fourier transform-based power spectral density, wavelet decomposition and ellipse-fitting morphology features described in more detail in Section \ref{sec:galaxy_zoo} and \ref{sec:ellipse}.

In addition to the feature extraction techniques, \astronomaly also has several post-processing methods, commonly used after features have been extracted, such as the dimensionality reduction technique principal component analysis \citep{pearson1901,hotelling1933} and feature scaling.

The choice of features is key because they will inevitably dictate what types of anomalies can be detected. As an example, in the Galaxy Zoo application below, we selected a feature extraction method that would be most sensitive to the shape of galaxies, thus successfully detecting unusual morphologies such as merging galaxies. However, ``green peas'' galaxies, a rare galaxy type studied extensively in the Galaxy Zoo dataset \citep{Cardamone2009}, would not be detected with these features as they have no colour sensitivity. The No Free Lunch Theorem \citep{Wolpert1996, Wolpert1997} dictates that it is not possible to design an algorithm that is guaranteed to detect all conceivable anomalies. A more intuitive way to think of this is that there would always be a way to design a pathological anomaly type that could not be detected by a particular algorithm. This can be mitigated by using combinations of feature extraction methods and multiple anomaly detection algorithms, but it is important to remember that no approach is ever guaranteed to be complete.

Deep learning methods offer an exciting alternative for many machine learning problems due to their ability to bypass the feature extraction phase. However there are few mature deep learning based anomaly detection approaches (although see \cite{sadr2019} for one promising algorithm, as well as \cite{polsterer2019}) and while some deep learning based methods were explored in creating \astronomaly and included in the code, the complexity, high computational cost and relatively low performance meant that we opted to leave a deep learning approach for future work.

\subsection{Anomaly detection}
\label{sec:anomaly_algorithms}
Machine learning is most commonly used for supervised learning where a training set is available. By definition, no or only very small training sets would be available for anomaly detection. Unsupervised machine learning algorithms, however, are able to run on unlabelled datasets. These algorithms usually work by learning what ``normal'' objects are and then assigning high anomaly scores to outliers that lie far from the normal population in feature space.

\astronomaly currently incorporates two anomaly detection algorithms, implemented with the \texttt{scikit\-learn} package \citep{pedregosa2011}, both of which are used in the applications in this work:
\begin{itemize}
 \item {\bfseries Isolation forest} - iForest \citep{liu2008} is an anomaly detection algorithm based on the popular classification algorithm random forests. Isolation forest works by attempting to isolate each object in the dataset using an ensemble of decision trees. At each node in the decision tree, a random feature and split value is chosen and the tree is grown until each data point is isolated. Intuitively, an anomaly is easier to isolate than a ``normal'' point that is very close (in feature space) to other points. Thus the decision path for an anomaly will be shorter and it will be assigned a higher anomaly score. iForest has high performance even for high dimensional datasets with many redundant features \citep{liu2008}.
 \item {\bfseries Local outlier factor} - LOF \citep{breunig2000} is a density-based anomaly detection algorithm. The local density in feature space is typically estimated by computing the ``reachability distance'' to an object's nearest neighbours. An object is determined to be an anomaly if it resides in a low density region. LOF is often particularly useful for datasets where anomalies lie fairly close to the ``normal'' class.
\end{itemize}
While these two algorithms both have good performance for the datasets we worked with, new anomaly detection algorithms can easily be incorporated into \astronomaly.

\subsection{Visualisation}
\label{sec:visualisation}
To more easily visualise the feature space and where anomalies are located therein, \astronomaly can apply mapping tools to embed high dimensional features into a low dimensional visualisation space. We currently only make use of t-Distributed Stochastic Neighbour Embedding \citep{vandermaaten2008}. t-SNE  is a visualisation tool that embeds a high dimensional feature space into a two or three dimensional visualisation space. It works by ensuring points that are close together in the original space remain nearby in the low dimensional space. t-SNE can help indicate at a glance what underlying structure may exist in the data and show possible clusters of similar objects. It should be noted that due to the stochastic nature of t-SNE (introduced to avoid being trapped in local minima), it can only be used for visualisation and not for dimensionality reduction. A newer technique, Uniform Manifold Approximation and Projection (UMAP; \citet{mcinnes2018}), offers a promising alternative which will be added in future. Figure \ref{fig:screenshot_visualisation} shows the web interface of \astronomaly with an interactive t-SNE plot.

\noindent
\begin{minipage}{1\linewidth}
    \gap
    \centering
    \includegraphics[width=1\linewidth]{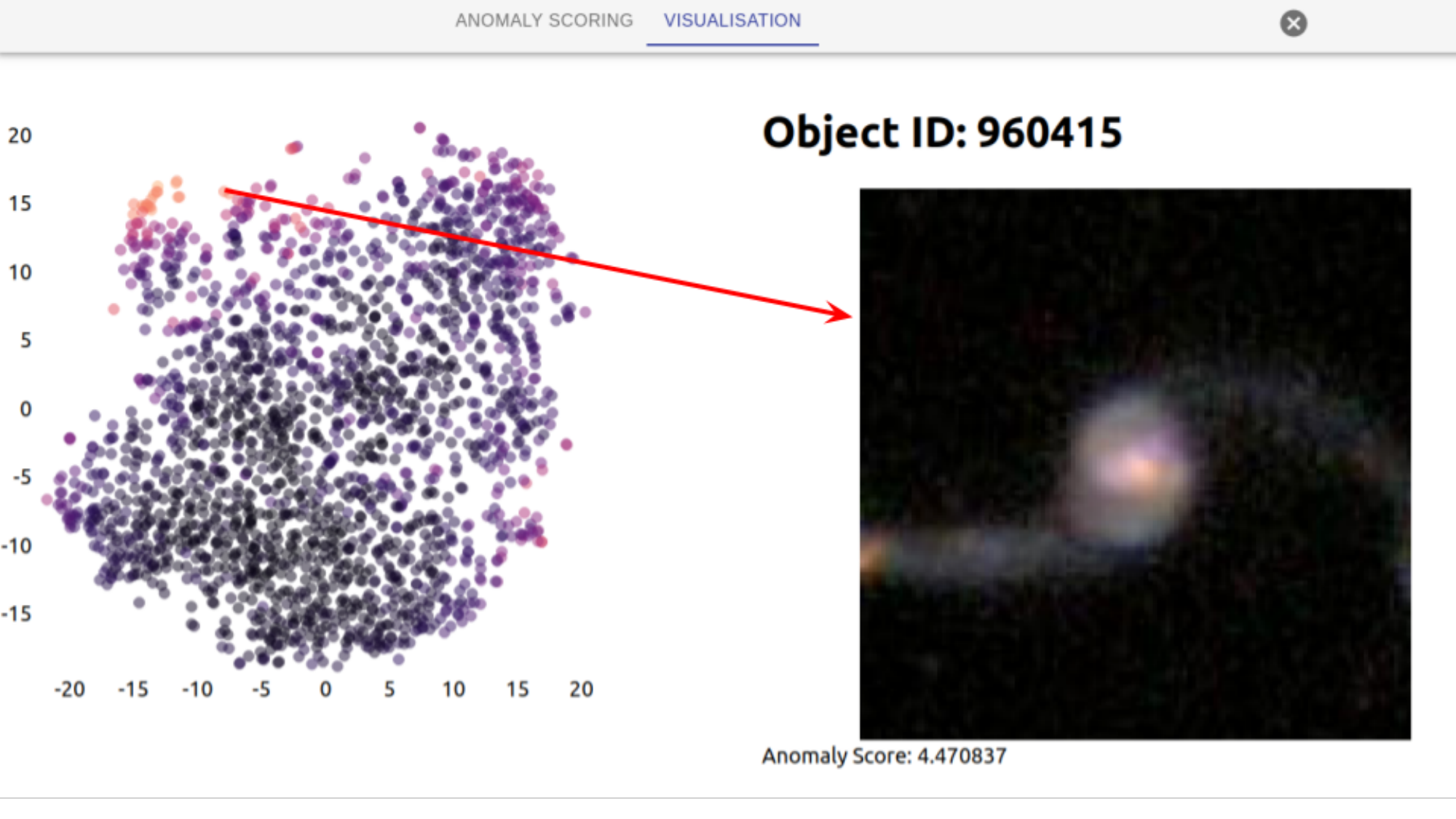}
    \captionof{figure}{A screenshot of the \astronomaly visual interface showing an interactive t-SNE plot, coloured by anomaly score (lighter colour indicates a more anomalous object). The t-SNE axes are in arbitrary units. Each point represents a different object in the dataset, which is displayed on the right by clicking on that point (visualised by an arrow).}
    \label{fig:screenshot_visualisation}
\end{minipage}

\subsection{Active learning applied to anomaly detection}

\astronomaly implements the active learning approach described in Section \ref{sec:active_learning}. The anomaly detection algorithm used is interchangeable, since all the active learning needs is the anomaly scores and user relevance scores. We developed a JavaScript frontend to allow users to rapidly input the relevance score, as well as view the data in an interactive manner. Active learning enables \astronomaly to act as a kind of ``recommendation engine'' for astronomical anomalies, where the resource of human viewing time is optimised by reducing the number of irrelevant anomalies seen by the user.

\section{Anomaly detection performance evaluation}
\label{sec:evaluation}
Any machine learning algorithm must be evaluated, but this is difficult in general for anomaly detection because by definition there is no training data in unsupervised learning. In addition, many commonly used metrics such as the area-under-curve of a ROC curve \citep{green1966, hanley1982, spackman1989} or the log loss \citep{bishop2006} tend not to be very differentiating in the case of highly unbalanced classes as they usually become dominated by the effects of the majority class.
However, other metrics can be useful particularly for simulated data or data which contains a sample of anomalies, to evaluate the performance of anomaly detection algorithms and our active learning approach. We caution the reader that \astronomaly remains a fundamentally subjective tool and the definition of ``anomaly'' will differ between users so all metrics should not be considered with finality.

We make use of two metrics to evaluate performance for data for which we have a ground truth:
\begin{itemize}
\item {\bfseries Rank Weighted Score (RWS)} - In anomaly detection (unlike in classification) the order of the output anomalies is important, given that we want to quickly present the human user with very interesting anomalies. We thus make use of the Rank Weighted Score (RWS), first described in \cite{roberts2019}, that assigns more value when an anomalous object is ranked more highly. The number of objects, $N$, must first be selected. This can often be naturally informed by how many objects a human may reasonably want to look at. In the case of simulated data or a labelled subset, $N$ can be set equal to the true number of anomalies. The RWS is defined as:
\begin{equation}
    \label{eq:rws}
S_{\rm RWS} = \frac{1}{S_0} \sum_{i=1}^N w_i I_i
\end{equation}
where the weights are:
\begin{equation}
w_i = (N + 1 - i)
\end{equation}
Additionally, $S_0 = N(N+1)/2$ and $I_i$ is an indicator variable: $I_i=1$ if object $i$ is an anomaly, and zero otherwise. Thus the RWS has a range of 0, if no anomalies were found, to 1 if all $N$ objects were correctly identified as anomalies.
\item {\bfseries Recall} - We also make use of the recall, which is simply how many anomalies have been detected after viewing $N$ objects. While this is a crude metric, it corresponds well with user experience where it is important that many anomalies are detected ``up front'' (early on in the ranked list).
\end{itemize}

\section{Application of \astronomaly to Galaxy Zoo data}
\label{sec:galaxy_zoo}
\subsection{Data description}
Galaxy Zoo \citep{lintott2008, lintott2011, willett2013} is a citizen science project which makes use of volunteers to label thousands of galaxies from a variety of data sources. The Galaxy Zoo data provides an excellent dataset of high signal-to-noise, resolved galaxies to test \astronomaly on.

We make use of the well-packaged version of the Galaxy Zoo 2 data on the public competition platform kaggle\footnote{\url{https://www.kaggle.com/c/galaxy-zoo-the-galaxy-challenge}}, which has a matching answer key described in \cite{willett2013zenodo}\footnote{\url{https://zenodo.org/record/3565489##.X2y6ypNKh-U}}. There are 61578 galaxies in the dataset, each represented by a 400x400 pixel cutout. The galaxies are classified by volunteers using a series of questions that follow a decision tree structure. Each galaxy is classified by at least 40 volunteers allowing each class to be assigned a probability. 

One of the questions asked of volunteers (corresponding to Class 6.1) ``Is there anything ``odd'' about the galaxy?'', can be useful as a ground truth for anomalies. There is a wide range of different types of objects that are considered ``odd'' by the Galaxy Zoo volunteers. It is therefore a very useful example of the broad range of possible objects that could be considered anomalous and the challenge of attempting to detect such wide diversity.

% \newpage
\noindent
\begin{minipage}{1\linewidth}
\begin{center}
\begin{minipage}{1\linewidth}
    \begin{minipage}{1\linewidth}
        \includegraphics[width=1\linewidth]{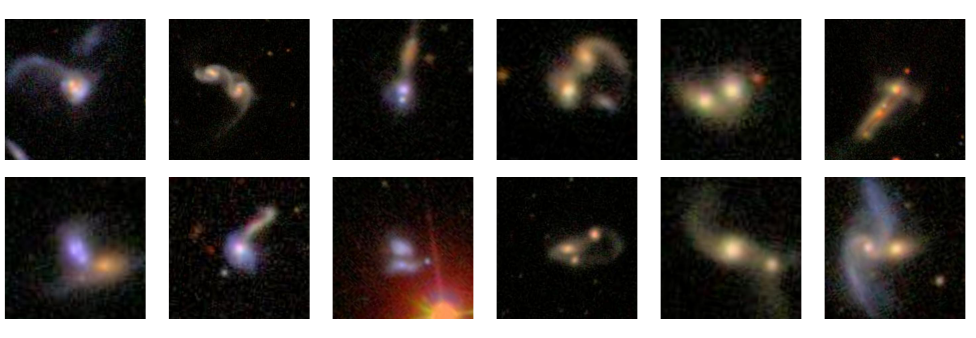}
        \captionof{subfigure}{\astronomaly: Top twelve most anomalous objects after applying active learning, showing how most artefacts are removed.}
        \label{fig:gal_2}
        \gap
    \end{minipage}
    \begin{minipage}{1\linewidth}
        \includegraphics[width=1\linewidth]{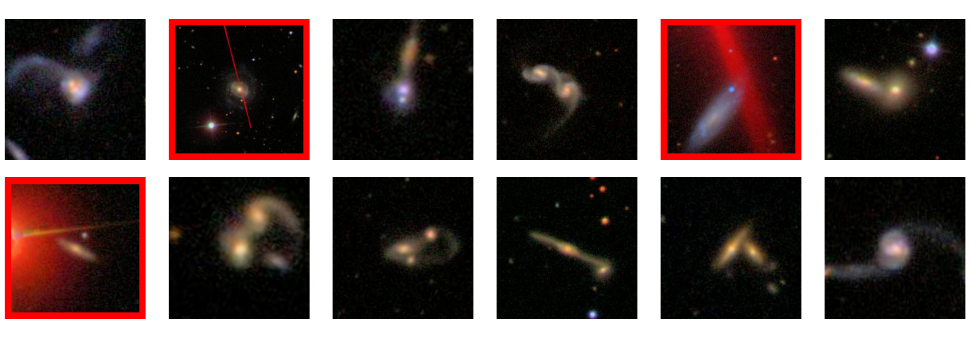}
        \captionof{subfigure}{No active learning: Top twelve most anomalous objects using isolation forest (artefacts are highlighted with a red border).}
        \label{fig:gal_1}
        \gap
    \end{minipage}
    
    \begin{minipage}{1\linewidth}
        \captionsetup[subfigure]{justification=raggedright, format=hang, singlelinecheck=false}
        \includegraphics[width=1\linewidth]{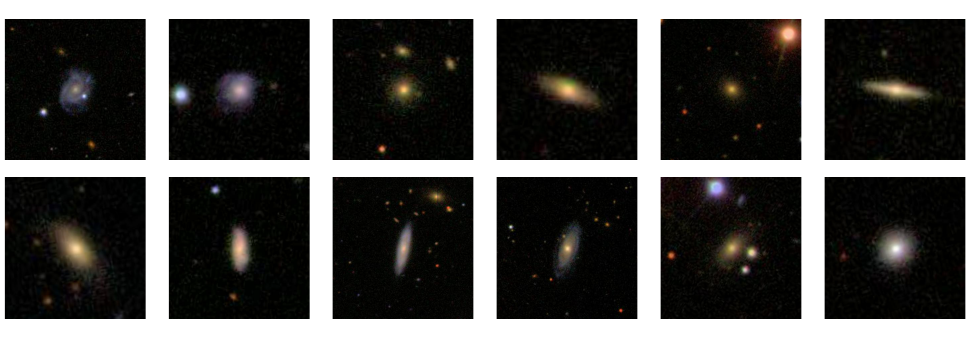}
        \captionof{subfigure}{Random examples.}
        \label{fig:gal_0}
        \gap
    \end{minipage}
    \end{minipage}
    \end{center}
    \captionof{figure}{Galaxy Zoo data examples using (a) \astronomaly with active learning to rank objects, (b) the iForest algorithm only to rank objects and (c) a random ordering for comparison. Notice that 
    \astronomaly's active learning removes most of the irrelevant artefacts.}
    \label{fig:galaxy_examples}
\end{minipage}

\subsection{Analysis}
We apply sigma-clipping to remove bright sources around the central galaxy that could otherwise distort results. We also develop a novel feature extraction technique based on ellipse fitting, which is conceptually simple although somewhat detailed in implementation. The essential idea is that morphologically ``boring'' galaxies tend to be well described by ellipses while a more complex structure will be poorly fit by an ellipse. We use contour finding at different brightness thresholds and fit an ellipse to each level, essentially using the various ellipse parameters as features. The full preprocessing and feature extraction procedure is described in \ref{sec:ellipse}.

We found these features were far more effective for this dataset than any others tried (including features based on moments, the FFT and deep autoencoders). They have the added advantage of being computationally extremely fast (see \ref{sec:computational} for a timing analysis), as they make use of the OpenCV library \citep{bradski2000} and being fairly robust to noise and varying sizes of sources. This method results in 21 features describing the ellipses at different threshold levels.

After we extract the features for the full dataset, we apply the anomaly detection algorithm isolation forest (iForest) to determine anomaly scores for each galaxy. We set \texttt{n\_estimators} to 100.
To test our active learning approach, we use the answers to the question ``is there something odd about this galaxy'' (Class 6.1) as a human labelled relevance score. The score is normalised from the input range of 0-1 (proportion of users that answered yes) to a range of 0-5 (integer) to emulate what would be used in the interface. Note that we perform no additional cuts on what is considered relevant so this should illustrate some of the ambiguity that may come from human labelling. We use the Class 6.1 scores to label the first 200 most anomalous (as scored by iForest) galaxies and then apply our active learning algorithm to predict a new relevance score for the remaining galaxies in the dataset.

\subsection{Results}
Figure \ref{fig:galaxy_examples} gives some illustrative examples of the Galaxy Zoo data. The bottom two rows are randomly drawn examples, demonstrating that most galaxies are ``boring'' ellipticals or spirals. The next two rows show the highest ranked anomalous objects according to iForest. It can be seen that with the features used, iForest is most sensitive to merging or interacting galaxies. Naturally there are a few artefacts which also result in high anomaly scores. The top two rows show the most anomalous objects after applying active learning, showing how the irrelevant objects have been downweighted allowing the user to focus on more interesting objects.

To further evaluate our results, we needed a dataset of firmly classified anomalies in order to compute the metrics described in Section \ref{sec:evaluation}. To do this, we extracted all objects with a Class 6.1 score greater than 0.9, which means at least 90\% of the volunteers labelled the galaxy as odd. Note that this is a pessimistic view, as there are surely many objects with a lower score that an expert would consider to be anomalous. However it seemed the least biased way to create a reliable ground truth anomaly dataset for testing purposes. This results in 924 anomalies (out of 61578 galaxies).

Figure \ref{fig:galaxy_metrics} shows on the left the recall (how many anomalies have been detected) as a function of index in a list of galaxies ranked by: random ordering, the iForest score and the trained active learning score. It can be seen that in particular, active learning improves the ranking early on in the list allowing the human user to prioritise the most interesting objects. \ml{This early phase where active learning provides significant benefits can be extended by providing more labelled training data.} The right panel of Figure \ref{fig:galaxy_metrics} illustrates the same point but using the rank weighted score (RWS) as a function of $N$ (the variable in the rank weighted score definition).

By making use of the Galaxy Zoo merger catalogue \citep{darg2010a, darg2010b}, we find that about two thirds of the anomalies detected by \astronomaly (in other words, ranked highly in the list) are galaxy mergers. \ml{60\% of the mergers listed in the catalogue are detected within the first 10\% of the ranked list of anomalies. The objects that are missed tend to be galaxies with multiple cores, indicating that the feature extractor can still be improved. We note that \astronomaly detects several objects that appear to be good candidates for mergers, yet were not ranked highly (more than 80\%) by the citizen scientists. This indicates that \astronomaly could be used to detect some objects missed by automated classifiers and citizen scientists.} The remainder of the anomalous objects tend to be either real compact groups of galaxies, \ml{galaxies with tidal tails} or chance alignments between galaxies or a galaxy and a nearby star. The algorithm can obviously be improved to reduce false positives, however it clearly does succeed in identifying objects that are in some way visually different.

Overall we find \astronomaly is able to successfully recover many of the interesting galaxies in the Galaxy Zoo dataset and is improved by making use of active learning.

\noindent
\begin{minipage}{1\linewidth}
    \begin{center}
    \begin{minipage}{1\linewidth}
        \begin{minipage}{0.5\linewidth}
               \includegraphics[width=1\linewidth]{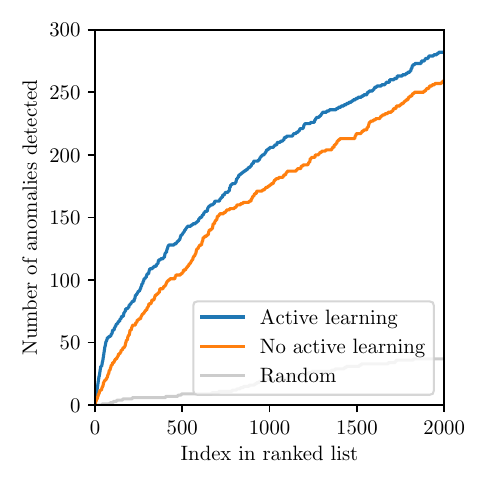}
               \captionof{subfigure}{}
               \label{fig:galaxy_cumulative}
               
        \end{minipage}
        \begin{minipage}{0.5\linewidth}
               \includegraphics[width=1\linewidth]{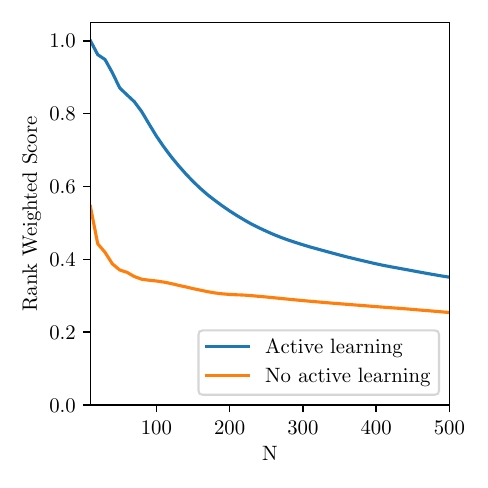}
               \captionof{subfigure}{}
               \label{fig:galaxy_rws}
               
        \end{minipage}
        \end{minipage}
    \end{center}
    \captionof{figure}{The effect of active learning on (a) the cumulative number of anomalies detected by that index and (b) the rank-weighted score for different values of $N$, the parameter in Equation \ref{eq:rws}. For scoring purposes, an anomaly is defined as an object that had a probability of greater than 0.9 of being labelled ``odd'' by a human. The effect of training is most pronounced for the highly ranked objects. While the false positive rate appears high, in practice many genuinely interesting galaxies do not make the 0.9 probability threshold to be considered anomalies here, so this should be seen as a pessimistic view.}
    \label{fig:galaxy_metrics}
    
\end{minipage}

\section{Application of \astronomaly to simulated data}
\label{sec:simulations}
\subsection{Description of simulated data}
To fully test \astronomaly, particularly our active learning approach, we need a dataset for which a ground truth is known. While the Galaxy Zoo dataset is labelled, these labels are subjective and correspond to many different classes of anomalies. We thus use an additional simulated dataset which, while contrived, thoroughly tests our approach.

We use the same simulations to those used in \cite{roberts2019}. These are meant to loosely emulate light curves but are not in any way realistic. We use simple one dimensional mathematical functions to simulate our data, each function belonging to a different class. We simulate two ``normal'' (majority) classes and three anomaly classes. Each function has parameters which, when generating the data, are randomly drawn from a Gaussian distribution, creating variability in the data. \ml{This includes varying the phase and amplitudes of the signals. We use 100 $x$-values, evenly spaced between zero and one. In the time-series analogy, this would correspond to data interpolated onto a regular grid over the same time period. Because the $x$-values are identical for all classes, they have no value for distinguishing anomalies and so only the $y$-values are used, meaning there are 100 features for each object.} 

The class functions, their corresponding parameter distributions and number of objects in the dataset are given in Table \ref{tab:functions}. \ml{The two inlier classes are created using a sine function and a parabola. The three outlier classes are made up of a step function, an exponential and a sum-of-sines function. Figure \ref{fig:sim_examples} shows some examples of the five different classes.} With three distinct (and known) classes of anomalies, we can now objectively test the active learning framework by choosing the relevance of each class and labelling them as a human would.

\begin{table}[!h]
\centering
\begin{tabular}{ccccc}
\hline
Class label & Type & Functional Form & Parameter distributions & Number of objects\\
\hline
0 & Inlier & $y = \rm{sin}(\omega x)$ & 
\begin{tabular}{@{}c@{}}
\\
 $ \omega \sim \mathcal{N}(5,2)$\\
 \\
\end{tabular}
& 25000\\
\hline
1 & Inlier & $y = \alpha x^2 + \beta x + \gamma$ & 
\begin{tabular}{@{}c@{}}

$\alpha \sim  \mathcal{N}(0.5, 0.2)$\\
$\beta \sim  \mathcal{N}(0.5, 0.2)$\\
$\gamma \sim  \mathcal{N}(0, 0.2)$\\

\end{tabular}
& 24500\\
\hline
2 & Outlier & $y =  h$ if  $x \leq x_0$, else $y=0$ &
\begin{tabular}{@{}c@{}}

$h \sim \mathcal{N}(1, 0.3)$\\
$x_0 \sim \mathcal{N}(0.5, 0.2)$\\
\\
\end{tabular} 
& 150\\
\hline
3 & Outlier & $y = A \, \exp \left( -\bigg(\frac{x-\mu}{w} \bigg)^2 \right)$ &
\begin{tabular}{@{}c@{}}
 $A  \sim  \mathcal{N}(0.5, 0.2)$\\
 $\mu  \sim  \mathcal{N}(0.1, 0.05)$\\
 $w  \sim  \mathcal{N}(1, 0.5)$\\
\end{tabular}
& 150\\
\hline
4 & Outlier & 
\begin{tabular}{@{}c@{}}
$y = 0.2\bigg(\sin(\omega_1 x) + \sin(\omega_2 x) + $\\
$\sin(\omega_3 x) + \sin(\omega_4 x)  + \sin(\omega_5 x) \bigg)$ 
\end{tabular}&
\begin{tabular}{@{}c@{}}
\\
 $\omega_i  \sim  \mathcal{N}(30, 20)$\\
\\
\end{tabular}
& 150\\
\hline
\end{tabular}
\caption{Description of the functions used to create the simulated data. 99\% of the objects in the dataset are of the type ``inlier'' and 1\% are ``outliers''. Each class has the corresponding functional form with  parameters drawn randomly for each instance from Gaussian distributions with hyperparameters specified in the table. \ml{Adapted from \cite{roberts2019}.}}
\label{tab:functions}
\end{table}

We generate about $50000$ curves of roughly equal number of objects from class 0 and 1 as training data. 
We add 1\% anomalies from classes 2, 3 and 4. For all classes, Gaussian noise with a standard deviation of $0.1$ is added.

\subsection{Analysis}
Because of the simplicity of the data, we use the raw data as features. We choose local outlier factor (LOF) as our algorithm for anomaly detection as we found it has slightly better performance than iForest (see Section \ref{sec:anomaly_algorithms} for a brief description). We set \texttt{n\_neighbors} to 100.

To test our active learning learning approach, we must decide which anomaly class is more relevant than the others. As this is arbitrary, we choose to decide that class 3 is incredibly boring (e.g. some instrumental artefact), class 2 is somewhat interesting and class 4 is very interesting. We then performed an experiment to analyse how our chosen performance metrics change as more and more data are labelled. The data are ranked in order of most to least anomalous (according to LOF), then the first 10 are labelled with their relevance score (0 for class 3, 3 for class 2 and 5 for class 4). The active learning algorithm is then trained and the data reordered according to the new trained anomaly score. This is then repeated in sets of 10 labelled objects up to 100 labels. This experiment simulates what may happen as one repeatedly trains the algorithm to improve it, although in practice it would be better to label at least 50 or 100 objects before initial training to ensure a suitably sized training set. For this simple problem though, performance gains can be seen after just 10 labels.

\subsection{Results}
The bottom two rows of Figure \ref{fig:sim_examples} shows six randomly drawn examples of the data, which (not surprisingly) are all from one of the two classes that make up 99\% of the data. The next two rows show the top six most anomalous objects as scored by LOF. We find that class 3 (neutral anomaly) tends to score particularly high while class 4 (our chosen most interesting class) is not seen as very anomalous, as it is more similar to the ``normal'' classes. After active learning is applied which downweights class 3 (boring class) and upweights class 4 (interesting class), we see in the top two rows that the interesting anomalies now appear high up in the rankings.

\noindent
\begin{minipage}{1\linewidth}
\begin{center}
\begin{minipage}{1\linewidth}
    
    \begin{minipage}{1\linewidth}
        \includegraphics[width=1\linewidth]{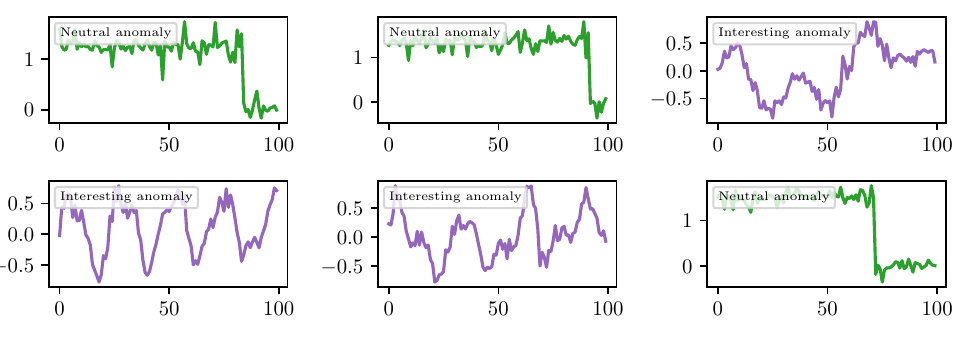}
        \captionof{subfigure}{\astronomaly: Top six most anomalous objects after applying active learning}
        \gap
    \end{minipage}
    \begin{minipage}{1\linewidth}
        \includegraphics[width=1\linewidth]{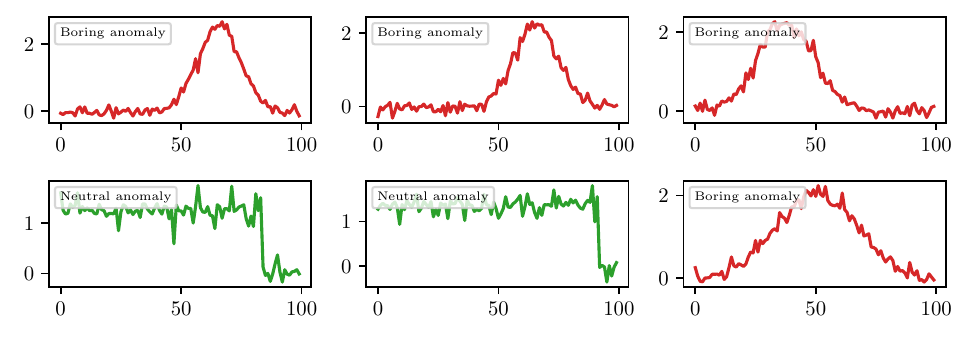}
        \captionof{subfigure}{No active learning: Top six most anomalous objects using LOF}
        \gap
    \end{minipage}
    \begin{minipage}{1\linewidth}
        \includegraphics[width=1\linewidth]{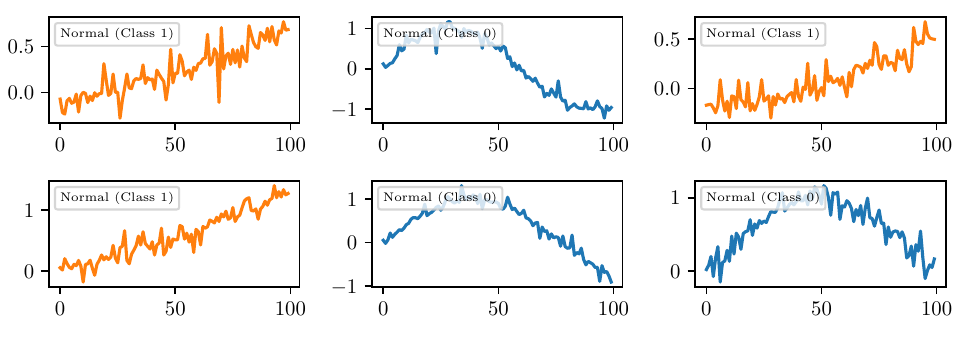}
        \captionof{subfigure}{Random examples}
    \end{minipage}
    \end{minipage}
    \end{center}
    \captionof{figure}{Simulated data examples using (a) active learning with human labels, (b) the LOF algorithm alone and (c) a random ordering for contrast.}
    \label{fig:sim_examples}
\end{minipage}

\noindent
\begin{minipage}{1\linewidth}
    \begin{center}
        \begin{minipage}{1\linewidth}
            \begin{minipage}{0.5\linewidth}
               \includegraphics[width=1\linewidth]{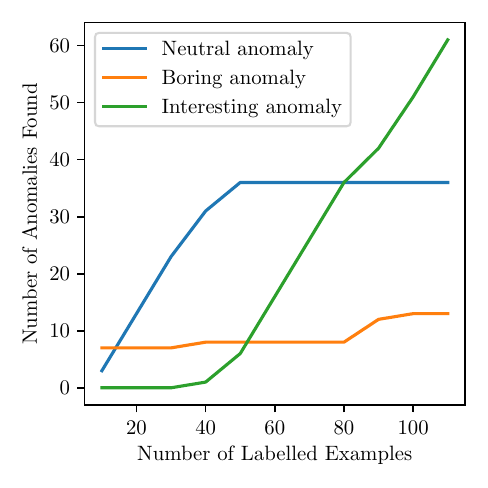}
               \captionof{subfigure}{}
            \end{minipage}
            \begin{minipage}{0.5\linewidth}
               \includegraphics[width=1\linewidth]{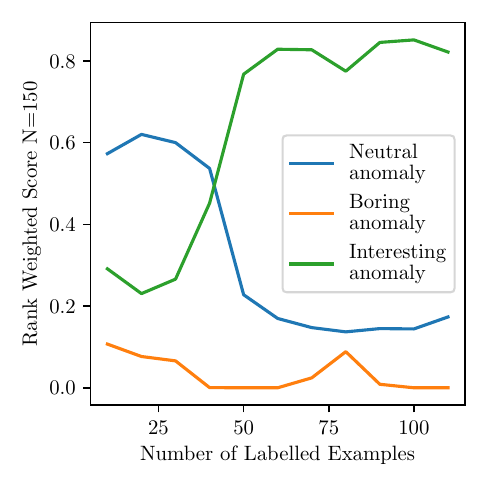}
                \captionof{subfigure}{}
            \end{minipage}
        \end{minipage}
    \end{center}
    \captionof{figure}{Active learning performance as a function of number of examples that have been seen and labelled. As the human scores data, the proportions of the anomaly classes changes among the most highly ranked objects. It can be seen in (a) that although class 2 (neutral) dominates early on, as more data are labelled class 4 (interesting) rapidly takes over, keeping class 3 (boring) ranked low, as the algorithm learns the user's preference. Similarly in (b), the RWS ($N=150$) climbs dramatically for class 4 as more data are labelled.}
    \label{fig:sim_training}
    \vspace{20pt}
\end{minipage}

The process of retraining the algorithm is illustrated in Figure \ref{fig:sim_training} where 10 examples are labelled, the algorithm is retrained, the data sorted by the new predicted anomaly score and then the next 10 are labelled. Initially the majority of the top rated anomalies are of class 2 (neutral) and 3 (boring), however after a few examples of class 4 (interesting) are labelled, the algorithm quickly learns to upweight these and they dominate the rankings. This effect is echoed in the rank weighted score (RWS) where $N$ is chosen to be 150: class 4 (interesting) initially performs poorly but quickly improves as the algorithm trains. Class 2 (neutral) gets somewhat penalised as class 4 dominates and class 3 (boring) remains at a low score. \ml{It takes around 30 labelled examples before the active learning starts to improve the RWS. This is likely because the random forest regressor that is used to predict the user relevance score requires a certain amount of training data before achieving sufficient accuracy. The amount of training data required will depend on the complexity of the problem: a more complex feature space where anomalies are not as easily separated out will require more training.}

The RWS for different values of $N$ before and after applying active learning is shown in Figure \ref{fig:sim_rws}. This is a slightly different experiment from previously in that it shows the effect of retraining just once with the top 100 most anomalous objects labelled (as opposed to iteratively labelling objects). Class 3, initially the top performing anomaly class, gets completely downweighted by having a relevance score of 0. Class 4's RWS improves dramatically while class 2, with a ``middle'' relevance score of 3, does not change significantly. 

Figure \ref{fig:sim_cumulative} shows the cumulative sum of anomalies (also known as recall) as a function of index in a ranked list (with 0 being the object with the highest anomaly score) both before (dashed line) and after (solid line) retraining with 100 human labelled objects. This again clearly shows that class 4 objects are pushed higher up the list while class 3 gets downweighted, and only starts increasing after almost all the class 4 anomalies have been found.

As a final illustration of how the active learning improves the relevance for this example data, we refer the reader back to Figure \ref{fig:sim_features} which shows a low-dimensional visualisation of feature space coloured by anomaly score, and indicating where the labelled objects appear in said feature space, both before and after retraining. Thus it can be seen that the algorithm works by learning which areas of feature space are more or less relevant to the user.

\noindent
\begin{minipage}{1\linewidth}
    \begin{center}
        \begin{minipage}{1\linewidth}
            \begin{minipage}{0.5\linewidth}
               \includegraphics[width=1\linewidth]{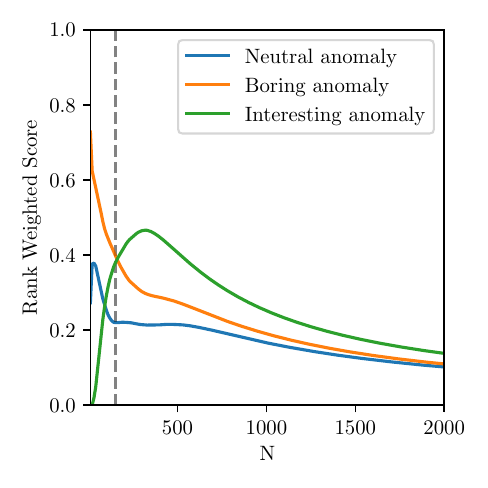}
               \captionof{subfigure}{}
            \end{minipage}
            \begin{minipage}{0.5\linewidth}
               \includegraphics[width=1\linewidth]{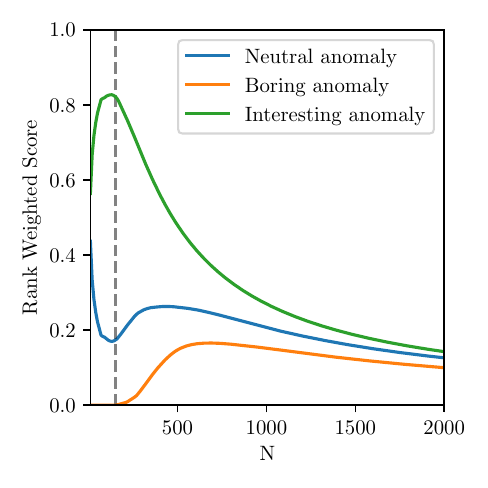}
                \captionof{subfigure}{}
            \end{minipage}
        \end{minipage}
    \end{center}
    \captionof{figure}{The change in the RWS as a function of $N$ (a) before and (b) after the top 100 most anomalous objects are labelled by a human and active learning is applied. It can be seen that before training, classes 3 (boring) and 2 (neutral) have a relatively high RWS. However, after these are downweighted by the human labelling and the scores adjusted, the RWS of class 4 (interesting) improves dramatically. A dashed line is used at $N=150$ to indicate the point where $N$ is equal to the number of anomalies in that class.}
    \label{fig:sim_rws}
\end{minipage}

\noindent
\begin{minipage}{1\linewidth}
    \begin{center}
        \begin{minipage}{1\linewidth}
               \includegraphics[width=1\linewidth]{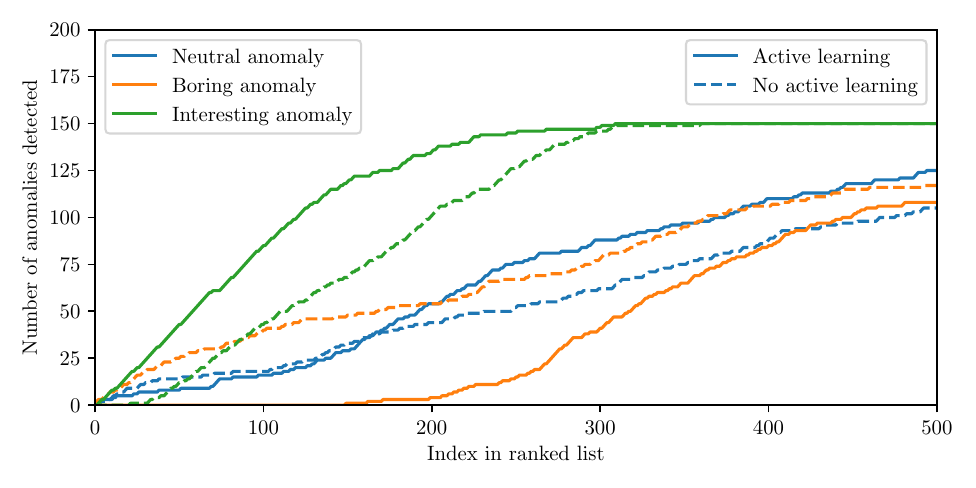}
        \end{minipage}
    \end{center}
    \captionof{figure}{The cumulative sum of anomalies in each class before and after active learning. This shows how the interesting class, class 4, dominates the most highly ranked objects after training while the ``boring'' class (in orange) is actively suppressed by active learning.}
    \label{fig:sim_cumulative}
\end{minipage}

\newpage
\section{Conclusions}
\label{sec:conclusions}
We have developed a general active learning anomaly detection framework, \astronomaly, to efficiently find interesting objects in large data volumes in astronomy.  \astronomaly is a publicly available package with a python backend, to run machine learning based anomaly detection, and a JavaScript frontend, to allow easy viewing and labelling of data. We have designed this framework to be applicable to almost any data type common in astronomy including images, one-dimensional data such as spectra and, in the near future, time series data. \astronomaly includes several feature extraction, preprocessing and machine learning techniques which are easily extendable.

In this work, we have focused on the insight that while machine learning may accurately identify anomalies in a dataset, not all anomalies are equally interesting. We argue that this concept of relevance is inherently subjective: what is of interest to one astronomer may not be of interest to another. We thus introduce a novel active learning approach that takes the relevance scores from a user and combines them with the raw anomaly score from machine learning. We incorporate a concept of uncertainty to ensure anomalies of a type never seen by a user are still ranked highly and thus not missed. The new retrained score can be used to more accurately rank objects according to user preference, effectively creating a recommendation engine from astronomical anomalies. 

We tested \astronomaly using Galaxy Zoo data, a large dataset of galaxy images labelled by citizen science volunteers. The Galaxy Zoo decision tree includes a question of ``is there anything odd about this galaxy?''. We make use of this label as our ground truth, finding that while there is large consensus of (for example) merging galaxies being anomalous, there is enough ambiguity in this score to illustrate the inherently subjective nature of anomaly labelling. The machine learning algorithm used (isolation forest), is able to accurately find many of these anomalies and rank them highly over the normal objects in the dataset. However it also detects many artefacts or chance alignments of galaxies as anomalies. We find that by applying our active learning approach, we can significantly reduce the number of these less interesting objects in the upper ranks of the sorted data. \ref{sec:computational} also shows that \astronomaly is extremely efficient, able to analyse thousands of objects quickly on a normal laptop. This means it will be able to scale effectively for datasets of millions of objects.

The novel feature extraction method we use, based on straightforward ellipse-fitting, is highly sensitive to unusual galaxy morphology such as mergers and tidal tails, but would be insensitive to galaxies with unusual colours, for example. This highlights the ``No Free Lunch'' theorem: in general, one cannot design an algorithm that can find every possible type of anomaly. Nonetheless, the approach used here was able to find a significant number of interesting objects and rank them highly, meaning a user would not have to search a large quantity of data to find many examples of anomalies. Our method is not guaranteed to be complete, however it does dramatically reduce time wasted searching through data when looking for new and unusual objects.

We further test our active learning approach with simulations that are meant to be visually similar to simple light curves and consist of several classes of anomalies. For this data, we have an (arbitrarily chosen) ground truth for the relevance of each type of anomaly and are able to show how the more relevant class is prioritised and ranked more highly than irrelevant anomalies, once active learning is performed. We find excellent performance (both in this case and for Galaxy Zoo) with just one to two hundred labels, which only takes a few minutes to apply using the \astronomaly web frontend. 

\astronomaly has also been used for light curve data \citep{webb2020} and this will be fully generalised and integrated into the package in the near future. Other future work will include extending the framework to operate in an online mode, allowing for continuous retraining. This is important for application to transient alert streams such as from the Zwicky Transient Facility \citep{bellm2018, graham2019} and the Vera C. Rubin Observatory. The Rubin Observatory is expected to detect around 10 million alerts every night during the Legacy Survey of Space and Time (LSST), which is a rich source of rare and interesting objects.

There is also scope for incorporating new feature extraction methods, anomaly detection algorithms and even new active learning approaches into the \astronomaly framework. We plan to include clustering techniques to allow rapid labelling of all data (not only anomalies) in an interactive environment. Additionally, we have so far only considered the case where a single user is interacting with \astronomaly. A very interesting use case would be where multiple users have access to the same data (each with their own preferences and objects of interests) and the results from multiple labels can be combined to create a meta-recommendation engine and recommend objects of interest based not just on the individuals preference, but that of other users as well. \ml{\astronomaly was built with a cloud-computing environment in mind: the backend can run on a remote server where large datasets are hosted while the frontend can be interacted with on any computer. While no paralellisation has been attempted in the initial release, the feature extraction (the most computationally intensive aspect) can be trivially parallelised. The anomaly detection algorithms used are expected to easily operate on millions of objects. The framework we have developed will thus allow \astronomaly to scale with the large datasets expected from telescopes such as the SKA and the Rubin Observatory.} 

\astronomaly is a versatile tool for scientific discovery in large astronomical datasets. By efficiently combining the automatic processing of machine learning with the intuition and knowledge of human scientists, \astronomaly aims to minimise laborious manual data exploration and maximise true scientific discovery.

\section*{Acknowledgements}
The authors would like to thank Ren\'{e}e Hlo\v{z}ek, Emille Ishida, Roy Maartens, Jason McEwen, Hiranya Peiris, and Mario Santos for comments on the draft and Nadeem Oozeer, Viral Parekh, Alireza Vafaei Sadr and Sara Webb for being early users of \astronomaly and providing feedback. We would also like to acknowledge the organisers and participants of the 2019 Kavli Summer Program in Astrophysics, where part of this work was done. The program was funded by the Kavli Foundation, the National Science Foundation, UC Santa Cruz and the Simons Foundation. Both authors acknowledge support from South African Radio Astronomy Observatory and the National Research Foundation (NRF) towards this  research. Opinions expressed and conclusions
arrived at, are those of the authors and are not  necessarily to be attributed to the NRF.

\biboptions{authoryear}
\bibliographystyle{cas-model2-names}
\bibliography{refs}

\begin{thebibliography}{50}
\expandafter\ifx\csname natexlab\endcsname\relax\def\natexlab#1{#1}\fi
\providecommand{\url}[1]{\texttt{#1}}
\providecommand{\href}[2]{#2}
\providecommand{\path}[1]{#1}
\providecommand{\DOIprefix}{doi:}
\providecommand{\ArXivprefix}{arXiv:}
\providecommand{\URLprefix}{URL: }
\providecommand{\Pubmedprefix}{pmid:}
\providecommand{\doi}[1]{\href{http://dx.doi.org/#1}{\path{#1}}}
\providecommand{\Pubmed}[1]{\href{pmid:#1}{\path{#1}}}
\providecommand{\bibinfo}[2]{#2}
\ifx\xfnm\relax \def\xfnm[#1]{\unskip,\space#1}\fi
%Type = Article
\bibitem[{{Astropy Collaboration} et~al.(2018){Astropy Collaboration},
  {Price-Whelan}, {Sip{H{o}}cz}, {G{"u}nther}, {Lim}, {Crawford}, {Conseil},
  {Shupe}, {Craig}, {Dencheva}, {Ginsburg}, {Vand erPlas}, {Bradley},
  {P{'e}rez-Su{'a}rez}, {de Val-Borro}, {Aldcroft}, {Cruz}, {Robitaille},
  {Tollerud}, {Ardelean}, {Babej}, {Bach}, {Bachetti}, {Bakanov}, {Bamford},
  {Barentsen}, {Barmby}, {Baumbach}, {Berry}, {Biscani}, {Boquien}, {Bostroem},
  {Bouma}, {Brammer}, {Bray}, {Breytenbach}, {Buddelmeijer}, {Burke},
  {Calderone}, {Cano Rodr{'i}guez}, {Cara}, {Cardoso}, {Cheedella}, {Copin},
  {Corrales}, {Crichton}, {D'Avella}, {Deil}, {Depagne}, {Dietrich}, {Donath},
  {Droettboom}, {Earl}, {Erben}, {Fabbro}, {Ferreira}, {Finethy}, {Fox},
  {Garrison}, {Gibbons}, {Goldstein}, {Gommers}, {Greco}, {Greenfield},
  {Groener}, {Grollier}, {Hagen}, {Hirst}, {Homeier}, {Horton}, {Hosseinzadeh},
  {Hu}, {Hunkeler}, {Ivezi{'c}}, {Jain}, {Jenness}, {Kanarek}, {Kendrew},
  {Kern}, {Kerzendorf}, {Khvalko}, {King}, {Kirkby}, {Kulkarni}, {Kumar},
  {Lee}, {Lenz}, {Littlefair}, {Ma}, {Macleod}, {Mastropietro}, {McCully},
  {Montagnac}, {Morris}, {Mueller}, {Mumford}, {Muna}, {Murphy}, {Nelson},
  {Nguyen}, {Ninan}, {N{"o}the}, {Ogaz}, {Oh}, {Parejko}, {Parley}, {Pascual},
  {Patil}, {Patil}, {Plunkett}, {Prochaska}, {Rastogi}, {Reddy Janga},
  {Sabater}, {Sakurikar}, {Seifert}, {Sherbert}, {Sherwood-Taylor}, {Shih},
  {Sick}, {Silbiger}, {Singanamalla}, {Singer}, {Sladen}, {Sooley},
  {Sornarajah}, {Streicher}, {Teuben}, {Thomas}, {Tremblay}, {Turner},
  {Terr{'o}n}, {van Kerkwijk}, {de la Vega}, {Watkins}, {Weaver}, {Whitmore},
  {Woillez}, {Zabalza} and {Astropy Contributors}}]{astropy2018}
\bibinfo{author}{{Astropy Collaboration}}, \bibinfo{author}{{Price-Whelan},
  A.M.}, \bibinfo{author}{{Sip{H{o}}cz}, B.M.}, \bibinfo{author}{{G{"u}nther},
  H.M.}, \bibinfo{author}{{Lim}, P.L.}, \bibinfo{author}{{Crawford}, S.M.},
  \bibinfo{author}{{Conseil}, S.}, \bibinfo{author}{{Shupe}, D.L.},
  \bibinfo{author}{{Craig}, M.W.}, \bibinfo{author}{{Dencheva}, N.},
  \bibinfo{author}{{Ginsburg}, A.}, \bibinfo{author}{{Vand erPlas}, J.T.},
  \bibinfo{author}{{Bradley}, L.D.}, \bibinfo{author}{{P{'e}rez-Su{'a}rez},
  D.}, \bibinfo{author}{{de Val-Borro}, M.}, \bibinfo{author}{{Aldcroft},
  T.L.}, \bibinfo{author}{{Cruz}, K.L.}, \bibinfo{author}{{Robitaille}, T.P.},
  \bibinfo{author}{{Tollerud}, E.J.}, \bibinfo{author}{{Ardelean}, C.},
  \bibinfo{author}{{Babej}, T.}, \bibinfo{author}{{Bach}, Y.P.},
  \bibinfo{author}{{Bachetti}, M.}, \bibinfo{author}{{Bakanov}, A.V.},
  \bibinfo{author}{{Bamford}, S.P.}, \bibinfo{author}{{Barentsen}, G.},
  \bibinfo{author}{{Barmby}, P.}, \bibinfo{author}{{Baumbach}, A.},
  \bibinfo{author}{{Berry}, K.L.}, \bibinfo{author}{{Biscani}, F.},
  \bibinfo{author}{{Boquien}, M.}, \bibinfo{author}{{Bostroem}, K.A.},
  \bibinfo{author}{{Bouma}, L.G.}, \bibinfo{author}{{Brammer}, G.B.},
  \bibinfo{author}{{Bray}, E.M.}, \bibinfo{author}{{Breytenbach}, H.},
  \bibinfo{author}{{Buddelmeijer}, H.}, \bibinfo{author}{{Burke}, D.J.},
  \bibinfo{author}{{Calderone}, G.}, \bibinfo{author}{{Cano Rodr{'i}guez},
  J.L.}, \bibinfo{author}{{Cara}, M.}, \bibinfo{author}{{Cardoso}, J.V.M.},
  \bibinfo{author}{{Cheedella}, S.}, \bibinfo{author}{{Copin}, Y.},
  \bibinfo{author}{{Corrales}, L.}, \bibinfo{author}{{Crichton}, D.},
  \bibinfo{author}{{D'Avella}, D.}, \bibinfo{author}{{Deil}, C.},
  \bibinfo{author}{{Depagne}, E.}, \bibinfo{author}{{Dietrich}, J.P.},
  \bibinfo{author}{{Donath}, A.}, \bibinfo{author}{{Droettboom}, M.},
  \bibinfo{author}{{Earl}, N.}, \bibinfo{author}{{Erben}, T.},
  \bibinfo{author}{{Fabbro}, S.}, \bibinfo{author}{{Ferreira}, L.A.},
  \bibinfo{author}{{Finethy}, T.}, \bibinfo{author}{{Fox}, R.T.},
  \bibinfo{author}{{Garrison}, L.H.}, \bibinfo{author}{{Gibbons}, S.L.J.},
  \bibinfo{author}{{Goldstein}, D.A.}, \bibinfo{author}{{Gommers}, R.},
  \bibinfo{author}{{Greco}, J.P.}, \bibinfo{author}{{Greenfield}, P.},
  \bibinfo{author}{{Groener}, A.M.}, \bibinfo{author}{{Grollier}, F.},
  \bibinfo{author}{{Hagen}, A.}, \bibinfo{author}{{Hirst}, P.},
  \bibinfo{author}{{Homeier}, D.}, \bibinfo{author}{{Horton}, A.J.},
  \bibinfo{author}{{Hosseinzadeh}, G.}, \bibinfo{author}{{Hu}, L.},
  \bibinfo{author}{{Hunkeler}, J.S.}, \bibinfo{author}{{Ivezi{'c}}, Z.},
  \bibinfo{author}{{Jain}, A.}, \bibinfo{author}{{Jenness}, T.},
  \bibinfo{author}{{Kanarek}, G.}, \bibinfo{author}{{Kendrew}, S.},
  \bibinfo{author}{{Kern}, N.S.}, \bibinfo{author}{{Kerzendorf}, W.E.},
  \bibinfo{author}{{Khvalko}, A.}, \bibinfo{author}{{King}, J.},
  \bibinfo{author}{{Kirkby}, D.}, \bibinfo{author}{{Kulkarni}, A.M.},
  \bibinfo{author}{{Kumar}, A.}, \bibinfo{author}{{Lee}, A.},
  \bibinfo{author}{{Lenz}, D.}, \bibinfo{author}{{Littlefair}, S.P.},
  \bibinfo{author}{{Ma}, Z.}, \bibinfo{author}{{Macleod}, D.M.},
  \bibinfo{author}{{Mastropietro}, M.}, \bibinfo{author}{{McCully}, C.},
  \bibinfo{author}{{Montagnac}, S.}, \bibinfo{author}{{Morris}, B.M.},
  \bibinfo{author}{{Mueller}, M.}, \bibinfo{author}{{Mumford}, S.J.},
  \bibinfo{author}{{Muna}, D.}, \bibinfo{author}{{Murphy}, N.A.},
  \bibinfo{author}{{Nelson}, S.}, \bibinfo{author}{{Nguyen}, G.H.},
  \bibinfo{author}{{Ninan}, J.P.}, \bibinfo{author}{{N{"o}the}, M.},
  \bibinfo{author}{{Ogaz}, S.}, \bibinfo{author}{{Oh}, S.},
  \bibinfo{author}{{Parejko}, J.K.}, \bibinfo{author}{{Parley}, N.},
  \bibinfo{author}{{Pascual}, S.}, \bibinfo{author}{{Patil}, R.},
  \bibinfo{author}{{Patil}, A.A.}, \bibinfo{author}{{Plunkett}, A.L.},
  \bibinfo{author}{{Prochaska}, J.X.}, \bibinfo{author}{{Rastogi}, T.},
  \bibinfo{author}{{Reddy Janga}, V.}, \bibinfo{author}{{Sabater}, J.},
  \bibinfo{author}{{Sakurikar}, P.}, \bibinfo{author}{{Seifert}, M.},
  \bibinfo{author}{{Sherbert}, L.E.}, \bibinfo{author}{{Sherwood-Taylor}, H.},
  \bibinfo{author}{{Shih}, A.Y.}, \bibinfo{author}{{Sick}, J.},
  \bibinfo{author}{{Silbiger}, M.T.}, \bibinfo{author}{{Singanamalla}, S.},
  \bibinfo{author}{{Singer}, L.P.}, \bibinfo{author}{{Sladen}, P.H.},
  \bibinfo{author}{{Sooley}, K.A.}, \bibinfo{author}{{Sornarajah}, S.},
  \bibinfo{author}{{Streicher}, O.}, \bibinfo{author}{{Teuben}, P.},
  \bibinfo{author}{{Thomas}, S.W.}, \bibinfo{author}{{Tremblay}, G.R.},
  \bibinfo{author}{{Turner}, J.E.H.}, \bibinfo{author}{{Terr{'o}n}, V.},
  \bibinfo{author}{{van Kerkwijk}, M.H.}, \bibinfo{author}{{de la Vega}, A.},
  \bibinfo{author}{{Watkins}, L.L.}, \bibinfo{author}{{Weaver}, B.A.},
  \bibinfo{author}{{Whitmore}, J.B.}, \bibinfo{author}{{Woillez}, J.},
  \bibinfo{author}{{Zabalza}, V.}, \bibinfo{author}{{Astropy Contributors}},
  \bibinfo{year}{2018}.
\newblock \bibinfo{title}{{The Astropy Project: Building an Open-science
  Project and Status of the v2.0 Core Package}}.
\newblock \bibinfo{journal}{aj} \bibinfo{volume}{156}, \bibinfo{pages}{123}.
\newblock \DOIprefix\doi{10.3847/1538-3881/aabc4f},
  \href{http://arxiv.org/abs/1801.02634}{\tt arXiv:1801.02634}.
%Type = Article
\bibitem[{{Astropy Collaboration} et~al.(2013){Astropy Collaboration},
  {Robitaille}, {Tollerud}, {Greenfield}, {Droettboom}, {Bray}, {Aldcroft},
  {Davis}, {Ginsburg}, {Price-Whelan}, {Kerzendorf}, {Conley}, {Crighton},
  {Barbary}, {Muna}, {Ferguson}, {Grollier}, {Parikh}, {Nair}, {Unther},
  {Deil}, {Woillez}, {Conseil}, {Kramer}, {Turner}, {Singer}, {Fox}, {Weaver},
  {Zabalza}, {Edwards}, {Azalee Bostroem}, {Burke}, {Casey}, {Crawford},
  {Dencheva}, {Ely}, {Jenness}, {Labrie}, {Lim}, {Pierfederici}, {Pontzen},
  {Ptak}, {Refsdal}, {Servillat} and {Streicher}}]{astropy2013}
\bibinfo{author}{{Astropy Collaboration}}, \bibinfo{author}{{Robitaille},
  T.P.}, \bibinfo{author}{{Tollerud}, E.J.}, \bibinfo{author}{{Greenfield},
  P.}, \bibinfo{author}{{Droettboom}, M.}, \bibinfo{author}{{Bray}, E.},
  \bibinfo{author}{{Aldcroft}, T.}, \bibinfo{author}{{Davis}, M.},
  \bibinfo{author}{{Ginsburg}, A.}, \bibinfo{author}{{Price-Whelan}, A.M.},
  \bibinfo{author}{{Kerzendorf}, W.E.}, \bibinfo{author}{{Conley}, A.},
  \bibinfo{author}{{Crighton}, N.}, \bibinfo{author}{{Barbary}, K.},
  \bibinfo{author}{{Muna}, D.}, \bibinfo{author}{{Ferguson}, H.},
  \bibinfo{author}{{Grollier}, F.}, \bibinfo{author}{{Parikh}, M.M.},
  \bibinfo{author}{{Nair}, P.H.}, \bibinfo{author}{{Unther}, H.M.},
  \bibinfo{author}{{Deil}, C.}, \bibinfo{author}{{Woillez}, J.},
  \bibinfo{author}{{Conseil}, S.}, \bibinfo{author}{{Kramer}, R.},
  \bibinfo{author}{{Turner}, J.E.H.}, \bibinfo{author}{{Singer}, L.},
  \bibinfo{author}{{Fox}, R.}, \bibinfo{author}{{Weaver}, B.A.},
  \bibinfo{author}{{Zabalza}, V.}, \bibinfo{author}{{Edwards}, Z.I.},
  \bibinfo{author}{{Azalee Bostroem}, K.}, \bibinfo{author}{{Burke}, D.J.},
  \bibinfo{author}{{Casey}, A.R.}, \bibinfo{author}{{Crawford}, S.M.},
  \bibinfo{author}{{Dencheva}, N.}, \bibinfo{author}{{Ely}, J.},
  \bibinfo{author}{{Jenness}, T.}, \bibinfo{author}{{Labrie}, K.},
  \bibinfo{author}{{Lim}, P.L.}, \bibinfo{author}{{Pierfederici}, F.},
  \bibinfo{author}{{Pontzen}, A.}, \bibinfo{author}{{Ptak}, A.},
  \bibinfo{author}{{Refsdal}, B.}, \bibinfo{author}{{Servillat}, M.},
  \bibinfo{author}{{Streicher}, O.}, \bibinfo{year}{2013}.
\newblock \bibinfo{title}{{Astropy: A community Python package for astronomy}}.
\newblock \bibinfo{journal}{\aap} \bibinfo{volume}{558}, \bibinfo{pages}{A33}.
\newblock \DOIprefix\doi{10.1051/0004-6361/201322068},
  \href{http://arxiv.org/abs/1307.6212}{\tt arXiv:1307.6212}.
%Type = Article
\bibitem[{Baron and Poznanski(2016)}]{Baron_2016}
\bibinfo{author}{Baron, D.}, \bibinfo{author}{Poznanski, D.},
  \bibinfo{year}{2016}.
\newblock \bibinfo{title}{The weirdest sdss galaxies: results from an outlier
  detection algorithm}.
\newblock \bibinfo{journal}{Monthly Notices of the Royal Astronomical Society}
  \bibinfo{volume}{465}, \bibinfo{pages}{4530–4555}.
\newblock \URLprefix \url{http://dx.doi.org/10.1093/mnras/stw3021},
  \DOIprefix\doi{10.1093/mnras/stw3021}.
%Type = Article
\bibitem[{Bellm et~al.(2018)Bellm, Kulkarni, Graham, Dekany, Smith, Riddle,
  Masci, Helou, Prince, Adams and et~al.}]{bellm2018}
\bibinfo{author}{Bellm, E.C.}, \bibinfo{author}{Kulkarni, S.R.},
  \bibinfo{author}{Graham, M.J.}, \bibinfo{author}{Dekany, R.},
  \bibinfo{author}{Smith, R.M.}, \bibinfo{author}{Riddle, R.},
  \bibinfo{author}{Masci, F.J.}, \bibinfo{author}{Helou, G.},
  \bibinfo{author}{Prince, T.A.}, \bibinfo{author}{Adams, S.M.},
  \bibinfo{author}{et~al.}, \bibinfo{year}{2018}.
\newblock \bibinfo{title}{The zwicky transient facility: System overview,
  performance, and first results}.
\newblock \bibinfo{journal}{Publications of the Astronomical Society of the
  Pacific} \bibinfo{volume}{131}, \bibinfo{pages}{018002}.
\newblock \URLprefix \url{http://dx.doi.org/10.1088/1538-3873/aaecbe},
  \DOIprefix\doi{10.1088/1538-3873/aaecbe}.
%Type = Book
\bibitem[{Bishop(2006)}]{bishop2006}
\bibinfo{author}{Bishop, C.M.}, \bibinfo{year}{2006}.
\newblock \bibinfo{title}{Pattern Recognition and Machine Learning (Information
  Science and Statistics)}.
\newblock \bibinfo{publisher}{Springer-Verlag}, \bibinfo{address}{Berlin,
  Heidelberg}.
%Type = Article
\bibitem[{Bradski(2000)}]{bradski2000}
\bibinfo{author}{Bradski, G.}, \bibinfo{year}{2000}.
\newblock \bibinfo{title}{{The OpenCV Library}}.
\newblock \bibinfo{journal}{Dr. Dobb's Journal of Software Tools} .
%Type = Article
\bibitem[{Breiman(2001)}]{breiman2001}
\bibinfo{author}{Breiman, L.}, \bibinfo{year}{2001}.
\newblock \bibinfo{title}{Random forests}.
\newblock \bibinfo{journal}{Machine Learning} \bibinfo{volume}{45},
  \bibinfo{pages}{5--32}.
\newblock \URLprefix \url{http://dx.doi.org/10.1023/A:1010933404324},
  \DOIprefix\doi{10.1023/A:1010933404324}.
%Type = Article
\bibitem[{Breunig et~al.(2000)Breunig, Kriegel, Ng and Sander}]{breunig2000}
\bibinfo{author}{Breunig, M.M.}, \bibinfo{author}{Kriegel, H.P.},
  \bibinfo{author}{Ng, R.T.}, \bibinfo{author}{Sander, J.},
  \bibinfo{year}{2000}.
\newblock \bibinfo{title}{Lof: Identifying density-based local outliers}.
\newblock \bibinfo{journal}{SIGMOD Rec.} \bibinfo{volume}{29},
  \bibinfo{pages}{93–104}.
\newblock \URLprefix \url{https://doi.org/10.1145/335191.335388},
  \DOIprefix\doi{10.1145/335191.335388}.
%Type = Article
\bibitem[{{Cardamone} et~al.(2009){Cardamone}, {Schawinski}, {Sarzi},
  {Bamford}, {Bennert}, {Urry}, {Lintott}, {Keel}, {Parejko}, {Nichol},
  {Thomas}, {Andreescu}, {Murray}, {Raddick}, {Slosar}, {Szalay} and
  {Vandenberg}}]{Cardamone2009}
\bibinfo{author}{{Cardamone}, C.}, \bibinfo{author}{{Schawinski}, K.},
  \bibinfo{author}{{Sarzi}, M.}, \bibinfo{author}{{Bamford}, S.P.},
  \bibinfo{author}{{Bennert}, N.}, \bibinfo{author}{{Urry}, C.M.},
  \bibinfo{author}{{Lintott}, C.}, \bibinfo{author}{{Keel}, W.C.},
  \bibinfo{author}{{Parejko}, J.}, \bibinfo{author}{{Nichol}, R.C.},
  \bibinfo{author}{{Thomas}, D.}, \bibinfo{author}{{Andreescu}, D.},
  \bibinfo{author}{{Murray}, P.}, \bibinfo{author}{{Raddick}, M.J.},
  \bibinfo{author}{{Slosar}, A.}, \bibinfo{author}{{Szalay}, A.},
  \bibinfo{author}{{Vandenberg}, J.}, \bibinfo{year}{2009}.
\newblock \bibinfo{title}{{Galaxy Zoo Green Peas: discovery of a class of
  compact extremely star-forming galaxies}}.
\newblock \bibinfo{journal}{\mnras} \bibinfo{volume}{399},
  \bibinfo{pages}{1191--1205}.
\newblock \DOIprefix\doi{10.1111/j.1365-2966.2009.15383.x},
  \href{http://arxiv.org/abs/0907.4155}{\tt arXiv:0907.4155}.
%Type = Article
\bibitem[{Clarke et~al.(2020)Clarke, Scaife, Greenhalgh and
  Griguta}]{Clarke2020}
\bibinfo{author}{Clarke, A.O.}, \bibinfo{author}{Scaife, A.M.M.},
  \bibinfo{author}{Greenhalgh, R.}, \bibinfo{author}{Griguta, V.},
  \bibinfo{year}{2020}.
\newblock \bibinfo{title}{Identifying galaxies, quasars, and stars with machine
  learning: A new catalogue of classifications for 111 million sdss sources
  without spectra}.
\newblock \bibinfo{journal}{Astronomy \& Astrophysics} \bibinfo{volume}{639},
  \bibinfo{pages}{A84}.
\newblock \URLprefix \url{http://dx.doi.org/10.1051/0004-6361/201936770},
  \DOIprefix\doi{10.1051/0004-6361/201936770}.
%Type = Article
\bibitem[{{Darg} et~al.(2010a){Darg}, {Kaviraj}, {Lintott}, {Schawinski},
  {Sarzi}, {Bamford}, {Silk}, {Andreescu}, {Murray}, {Nichol}, {Raddick},
  {Slosar}, {Szalay}, {Thomas} and {Vandenberg}}]{darg2010b}
\bibinfo{author}{{Darg}, D.W.}, \bibinfo{author}{{Kaviraj}, S.},
  \bibinfo{author}{{Lintott}, C.J.}, \bibinfo{author}{{Schawinski}, K.},
  \bibinfo{author}{{Sarzi}, M.}, \bibinfo{author}{{Bamford}, S.},
  \bibinfo{author}{{Silk}, J.}, \bibinfo{author}{{Andreescu}, D.},
  \bibinfo{author}{{Murray}, P.}, \bibinfo{author}{{Nichol}, R.C.},
  \bibinfo{author}{{Raddick}, M.J.}, \bibinfo{author}{{Slosar}, A.},
  \bibinfo{author}{{Szalay}, A.S.}, \bibinfo{author}{{Thomas}, D.},
  \bibinfo{author}{{Vandenberg}, J.}, \bibinfo{year}{2010}a.
\newblock \bibinfo{title}{{Galaxy Zoo: the properties of merging galaxies in
  the nearby Universe - local environments, colours, masses, star formation
  rates and AGN activity}}.
\newblock \bibinfo{journal}{\mnras} \bibinfo{volume}{401},
  \bibinfo{pages}{1552--1563}.
\newblock \DOIprefix\doi{10.1111/j.1365-2966.2009.15786.x},
  \href{http://arxiv.org/abs/0903.5057}{\tt arXiv:0903.5057}.
%Type = Article
\bibitem[{{Darg} et~al.(2010b){Darg}, {Kaviraj}, {Lintott}, {Schawinski},
  {Sarzi}, {Bamford}, {Silk}, {Proctor}, {Andreescu}, {Murray}, {Nichol},
  {Raddick}, {Slosar}, {Szalay}, {Thomas} and {Vandenberg}}]{darg2010a}
\bibinfo{author}{{Darg}, D.W.}, \bibinfo{author}{{Kaviraj}, S.},
  \bibinfo{author}{{Lintott}, C.J.}, \bibinfo{author}{{Schawinski}, K.},
  \bibinfo{author}{{Sarzi}, M.}, \bibinfo{author}{{Bamford}, S.},
  \bibinfo{author}{{Silk}, J.}, \bibinfo{author}{{Proctor}, R.},
  \bibinfo{author}{{Andreescu}, D.}, \bibinfo{author}{{Murray}, P.},
  \bibinfo{author}{{Nichol}, R.C.}, \bibinfo{author}{{Raddick}, M.J.},
  \bibinfo{author}{{Slosar}, A.}, \bibinfo{author}{{Szalay}, A.S.},
  \bibinfo{author}{{Thomas}, D.}, \bibinfo{author}{{Vandenberg}, J.},
  \bibinfo{year}{2010}b.
\newblock \bibinfo{title}{{Galaxy Zoo: the fraction of merging galaxies in the
  SDSS and their morphologies}}.
\newblock \bibinfo{journal}{\mnras} \bibinfo{volume}{401},
  \bibinfo{pages}{1043--1056}.
\newblock \DOIprefix\doi{10.1111/j.1365-2966.2009.15686.x},
  \href{http://arxiv.org/abs/0903.4937}{\tt arXiv:0903.4937}.
%Type = Misc
\bibitem[{Doorenbos et~al.(2020)Doorenbos, Cavuoti, Brescia, D'Isanto and
  Longo}]{doorenbos2020}
\bibinfo{author}{Doorenbos, L.}, \bibinfo{author}{Cavuoti, S.},
  \bibinfo{author}{Brescia, M.}, \bibinfo{author}{D'Isanto, A.},
  \bibinfo{author}{Longo, G.}, \bibinfo{year}{2020}.
\newblock \bibinfo{title}{Comparison of outlier detection methods on
  astronomical image data}.
\newblock \href{http://arxiv.org/abs/2006.08238}{\tt arXiv:2006.08238}.
%Type = Misc
\bibitem[{Fluke et~al.(2020)Fluke, Hegarty and MacMahon}]{fluke2020}
\bibinfo{author}{Fluke, C.J.}, \bibinfo{author}{Hegarty, S.E.},
  \bibinfo{author}{MacMahon, C.O.M.}, \bibinfo{year}{2020}.
\newblock \bibinfo{title}{Understanding the human in the design of cyber-human
  discovery systems for data-driven astronomy}.
\newblock \href{http://arxiv.org/abs/2008.13112}{\tt arXiv:2008.13112}.
%Type = Article
\bibitem[{Giles and Walkowicz(2018)}]{Giles_2018}
\bibinfo{author}{Giles, D.}, \bibinfo{author}{Walkowicz, L.},
  \bibinfo{year}{2018}.
\newblock \bibinfo{title}{Systematic serendipity: a test of unsupervised
  machine learning as a method for anomaly detection}.
\newblock \bibinfo{journal}{Monthly Notices of the Royal Astronomical Society}
  \bibinfo{volume}{484}, \bibinfo{pages}{834–849}.
\newblock \URLprefix \url{http://dx.doi.org/10.1093/mnras/sty3461},
  \DOIprefix\doi{10.1093/mnras/sty3461}.
%Type = Article
\bibitem[{Graham et~al.(2019)Graham, Kulkarni, Bellm, Adams, Barbarino,
  Blagorodnova, Bodewits, Bolin, Brady, Cenko and et~al.}]{graham2019}
\bibinfo{author}{Graham, M.J.}, \bibinfo{author}{Kulkarni, S.R.},
  \bibinfo{author}{Bellm, E.C.}, \bibinfo{author}{Adams, S.M.},
  \bibinfo{author}{Barbarino, C.}, \bibinfo{author}{Blagorodnova, N.},
  \bibinfo{author}{Bodewits, D.}, \bibinfo{author}{Bolin, B.},
  \bibinfo{author}{Brady, P.R.}, \bibinfo{author}{Cenko, S.B.},
  \bibinfo{author}{et~al.}, \bibinfo{year}{2019}.
\newblock \bibinfo{title}{The zwicky transient facility: Science objectives}.
\newblock \bibinfo{journal}{Publications of the Astronomical Society of the
  Pacific} \bibinfo{volume}{131}, \bibinfo{pages}{078001}.
\newblock \URLprefix \url{http://dx.doi.org/10.1088/1538-3873/ab006c},
  \DOIprefix\doi{10.1088/1538-3873/ab006c}.
%Type = Article
\bibitem[{Green and Swets(1966)}]{green1966}
\bibinfo{author}{Green, D.}, \bibinfo{author}{Swets, J.}, \bibinfo{year}{1966}.
\newblock \bibinfo{title}{Signal detection theory and psychophysics}.
\newblock \bibinfo{journal}{Society} \bibinfo{volume}{1}, \bibinfo{pages}{521}.
%Type = Article
\bibitem[{Hanley and McNeil(1982)}]{hanley1982}
\bibinfo{author}{Hanley, J.A.}, \bibinfo{author}{McNeil, B.J.},
  \bibinfo{year}{1982}.
\newblock \bibinfo{title}{The meaning and use of the area under a receiver
  operating characteristic (roc) curve.}
\newblock \bibinfo{journal}{Radiology} \bibinfo{volume}{143},
  \bibinfo{pages}{29--36}.
%Type = Article
\bibitem[{Hocking et~al.(2017)Hocking, Geach, Sun and Davey}]{Hocking_2017}
\bibinfo{author}{Hocking, A.}, \bibinfo{author}{Geach, J.E.},
  \bibinfo{author}{Sun, Y.}, \bibinfo{author}{Davey, N.}, \bibinfo{year}{2017}.
\newblock \bibinfo{title}{An automatic taxonomy of galaxy morphology using
  unsupervised machine learning}.
\newblock \bibinfo{journal}{Monthly Notices of the Royal Astronomical Society}
  \bibinfo{volume}{473}, \bibinfo{pages}{1108–1129}.
\newblock \URLprefix \url{http://dx.doi.org/10.1093/mnras/stx2351},
  \DOIprefix\doi{10.1093/mnras/stx2351}.
%Type = Article
\bibitem[{Hotelling(1933)}]{hotelling1933}
\bibinfo{author}{Hotelling, H.}, \bibinfo{year}{1933}.
\newblock \bibinfo{title}{Analysis of a complex of statistical variables into
  principal components.}
\newblock \bibinfo{journal}{Journal of educational psychology}
  \bibinfo{volume}{24}, \bibinfo{pages}{417}.
%Type = Misc
\bibitem[{Ishida et~al.(2020)Ishida, Kornilov, Malanchev, Pruzhinskaya,
  Volnova, Korolev, Mondon, Sreejith, Malancheva and Das}]{ishida2020}
\bibinfo{author}{Ishida, E.E.O.}, \bibinfo{author}{Kornilov, M.V.},
  \bibinfo{author}{Malanchev, K.L.}, \bibinfo{author}{Pruzhinskaya, M.V.},
  \bibinfo{author}{Volnova, A.A.}, \bibinfo{author}{Korolev, V.S.},
  \bibinfo{author}{Mondon, F.}, \bibinfo{author}{Sreejith, S.},
  \bibinfo{author}{Malancheva, A.}, \bibinfo{author}{Das, S.},
  \bibinfo{year}{2020}.
\newblock \bibinfo{title}{Active anomaly detection for time-domain
  discoveries}.
\newblock \href{http://arxiv.org/abs/1909.13260}{\tt arXiv:1909.13260}.
%Type = Article
\bibitem[{Ivezi\'{c} et~al.(2019)Ivezi\'{c}, Kahn, Tyson, Abel, Acosta,
  Allsman, Alonso, AlSayyad, Anderson, Andrew and et~al.}]{Ivezic2019}
\bibinfo{author}{Ivezi\'{c}, Z.}, \bibinfo{author}{Kahn, S.M.},
  \bibinfo{author}{Tyson, J.A.}, \bibinfo{author}{Abel, B.},
  \bibinfo{author}{Acosta, E.}, \bibinfo{author}{Allsman, R.},
  \bibinfo{author}{Alonso, D.}, \bibinfo{author}{AlSayyad, Y.},
  \bibinfo{author}{Anderson, S.F.}, \bibinfo{author}{Andrew, J.},
  \bibinfo{author}{et~al.}, \bibinfo{year}{2019}.
\newblock \bibinfo{title}{Lsst: From science drivers to reference design and
  anticipated data products}.
\newblock \bibinfo{journal}{The Astrophysical Journal} \bibinfo{volume}{873},
  \bibinfo{pages}{111}.
\newblock \URLprefix \url{http://dx.doi.org/10.3847/1538-4357/ab042c},
  \DOIprefix\doi{10.3847/1538-4357/ab042c}.
%Type = Misc
\bibitem[{Kong et~al.(2020)Kong, Chen, Chen, Bhatia and Callot}]{kong2020}
\bibinfo{author}{Kong, L.}, \bibinfo{author}{Chen, L.}, \bibinfo{author}{Chen,
  M.}, \bibinfo{author}{Bhatia, P.}, \bibinfo{author}{Callot, L.},
  \bibinfo{year}{2020}.
\newblock \bibinfo{title}{Improve black-box sequential anomaly detector
  relevancy with limited user feedback}.
\newblock \href{http://arxiv.org/abs/2009.07241}{\tt arXiv:2009.07241}.
%Type = Article
\bibitem[{{Lintott} et~al.(2011){Lintott}, {Schawinski}, {Bamford}, {Slosar},
  {Land}, {Thomas}, {Edmondson}, {Masters}, {Nichol}, {Raddick}, {Szalay},
  {Andreescu}, {Murray} and {Vandenberg}}]{lintott2011}
\bibinfo{author}{{Lintott}, C.}, \bibinfo{author}{{Schawinski}, K.},
  \bibinfo{author}{{Bamford}, S.}, \bibinfo{author}{{Slosar}, A.},
  \bibinfo{author}{{Land}, K.}, \bibinfo{author}{{Thomas}, D.},
  \bibinfo{author}{{Edmondson}, E.}, \bibinfo{author}{{Masters}, K.},
  \bibinfo{author}{{Nichol}, R.C.}, \bibinfo{author}{{Raddick}, M.J.},
  \bibinfo{author}{{Szalay}, A.}, \bibinfo{author}{{Andreescu}, D.},
  \bibinfo{author}{{Murray}, P.}, \bibinfo{author}{{Vandenberg}, J.},
  \bibinfo{year}{2011}.
\newblock \bibinfo{title}{{Galaxy Zoo 1: data release of morphological
  classifications for nearly 900 000 galaxies}}.
\newblock \bibinfo{journal}{\mnras} \bibinfo{volume}{410},
  \bibinfo{pages}{166--178}.
\newblock \DOIprefix\doi{10.1111/j.1365-2966.2010.17432.x},
  \href{http://arxiv.org/abs/1007.3265}{\tt arXiv:1007.3265}.
%Type = Article
\bibitem[{{Lintott} et~al.(2008){Lintott}, {Schawinski}, {Slosar}, {Land},
  {Bamford}, {Thomas}, {Raddick}, {Nichol}, {Szalay}, {Andreescu}, {Murray} and
  {Vandenberg}}]{lintott2008}
\bibinfo{author}{{Lintott}, C.J.}, \bibinfo{author}{{Schawinski}, K.},
  \bibinfo{author}{{Slosar}, A.}, \bibinfo{author}{{Land}, K.},
  \bibinfo{author}{{Bamford}, S.}, \bibinfo{author}{{Thomas}, D.},
  \bibinfo{author}{{Raddick}, M.J.}, \bibinfo{author}{{Nichol}, R.C.},
  \bibinfo{author}{{Szalay}, A.}, \bibinfo{author}{{Andreescu}, D.},
  \bibinfo{author}{{Murray}, P.}, \bibinfo{author}{{Vandenberg}, J.},
  \bibinfo{year}{2008}.
\newblock \bibinfo{title}{{Galaxy Zoo: morphologies derived from visual
  inspection of galaxies from the Sloan Digital Sky Survey}}.
\newblock \bibinfo{journal}{\mnras} \bibinfo{volume}{389},
  \bibinfo{pages}{1179--1189}.
\newblock \DOIprefix\doi{10.1111/j.1365-2966.2008.13689.x},
  \href{http://arxiv.org/abs/0804.4483}{\tt arXiv:0804.4483}.
%Type = Inproceedings
\bibitem[{{Liu} et~al.(2008){Liu}, {Ting} and {Zhou}}]{liu2008}
\bibinfo{author}{{Liu}, F.T.}, \bibinfo{author}{{Ting}, K.M.},
  \bibinfo{author}{{Zhou}, Z.}, \bibinfo{year}{2008}.
\newblock \bibinfo{title}{Isolation forest}, in: \bibinfo{booktitle}{2008
  Eighth IEEE International Conference on Data Mining}, pp.
  \bibinfo{pages}{413--422}.
\newblock \DOIprefix\doi{10.1109/ICDM.2008.17}.
%Type = Article
\bibitem[{Van~der Maaten and Hinton(2008)}]{vandermaaten2008}
\bibinfo{author}{Van~der Maaten, L.}, \bibinfo{author}{Hinton, G.},
  \bibinfo{year}{2008}.
\newblock \bibinfo{title}{Visualizing data using t-sne}.
\newblock \bibinfo{journal}{Journal of Machine Learning Research}
  \bibinfo{volume}{9}, \bibinfo{pages}{85}.
%Type = Article
\bibitem[{{Maneewongvatana} and {Mount}(1999)}]{maneewongvatana1999}
\bibinfo{author}{{Maneewongvatana}, S.}, \bibinfo{author}{{Mount}, D.M.},
  \bibinfo{year}{1999}.
\newblock \bibinfo{title}{{Analysis of approximate nearest neighbor searching
  with clustered point sets}}.
\newblock \bibinfo{journal}{arXiv e-prints} ,
  \bibinfo{pages}{cs/9901013}\href{http://arxiv.org/abs/cs/9901013}{\tt
  arXiv:cs/9901013}.
%Type = Article
\bibitem[{Margalef-Bentabol et~al.(2020)Margalef-Bentabol, Huertas-Company,
  Charnock, Margalef-Bentabol, Bernardi, Dubois, Storey-Fisher and
  Zanisi}]{Margalef_Bentabol_2020}
\bibinfo{author}{Margalef-Bentabol, B.}, \bibinfo{author}{Huertas-Company, M.},
  \bibinfo{author}{Charnock, T.}, \bibinfo{author}{Margalef-Bentabol, C.},
  \bibinfo{author}{Bernardi, M.}, \bibinfo{author}{Dubois, Y.},
  \bibinfo{author}{Storey-Fisher, K.}, \bibinfo{author}{Zanisi, L.},
  \bibinfo{year}{2020}.
\newblock \bibinfo{title}{Detecting outliers in astronomical images with deep
  generative networks}.
\newblock \bibinfo{journal}{Monthly Notices of the Royal Astronomical Society}
  \bibinfo{volume}{496}, \bibinfo{pages}{2346–2361}.
\newblock \URLprefix \url{http://dx.doi.org/10.1093/mnras/staa1647},
  \DOIprefix\doi{10.1093/mnras/staa1647}.
%Type = Article
\bibitem[{Marianer et~al.(2020)Marianer, Poznanski and
  Prochaska}]{Marianer2020}
\bibinfo{author}{Marianer, T.}, \bibinfo{author}{Poznanski, D.},
  \bibinfo{author}{Prochaska, J.X.}, \bibinfo{year}{2020}.
\newblock \bibinfo{title}{A semisupervised machine learning search for
  never-seen gravitational-wave sources}.
\newblock \bibinfo{journal}{Monthly Notices of the Royal Astronomical Society}
  \bibinfo{volume}{500}, \bibinfo{pages}{5408–5419}.
\newblock \URLprefix \url{http://dx.doi.org/10.1093/mnras/staa3550},
  \DOIprefix\doi{10.1093/mnras/staa3550}.
%Type = Misc
\bibitem[{Martínez-Galarza et~al.(2020)Martínez-Galarza, Bianco, Crake,
  Tirumala, Mahabal, Graham and Giles}]{martinezgalarza2020}
\bibinfo{author}{Martínez-Galarza, J.R.}, \bibinfo{author}{Bianco, F.},
  \bibinfo{author}{Crake, D.}, \bibinfo{author}{Tirumala, K.},
  \bibinfo{author}{Mahabal, A.A.}, \bibinfo{author}{Graham, M.J.},
  \bibinfo{author}{Giles, D.}, \bibinfo{year}{2020}.
\newblock \bibinfo{title}{Where is waldo (and his friends)? a comparison of
  anomaly detection algorithms for time-domain astronomy}.
\newblock \href{http://arxiv.org/abs/2009.06760}{\tt arXiv:2009.06760}.
%Type = Article
\bibitem[{McInnes et~al.(2018)McInnes, Healy, Saul and
  Grossberger}]{mcinnes2018}
\bibinfo{author}{McInnes, L.}, \bibinfo{author}{Healy, J.},
  \bibinfo{author}{Saul, N.}, \bibinfo{author}{Grossberger, L.},
  \bibinfo{year}{2018}.
\newblock \bibinfo{title}{Umap: Uniform manifold approximation and projection}.
\newblock \bibinfo{journal}{Journal of Open Source Software}
  \bibinfo{volume}{3}, \bibinfo{pages}{861}.
\newblock \DOIprefix\doi{10.21105/joss.00861}.
%Type = Article
\bibitem[{Pearson(1901)}]{pearson1901}
\bibinfo{author}{Pearson, K.}, \bibinfo{year}{1901}.
\newblock \bibinfo{title}{Liii. on lines and planes of closest fit to systems
  of points in space}.
\newblock \bibinfo{journal}{The London, Edinburgh, and Dublin Philosophical
  Magazine and Journal of Science} \bibinfo{volume}{2},
  \bibinfo{pages}{559--572}.
%Type = Article
\bibitem[{Pedregosa et~al.(2011)Pedregosa, Varoquaux, Gramfort, Michel,
  Thirion, Grisel, Blondel, Prettenhofer, Weiss, Dubourg, Vanderplas, Passos,
  Cournapeau, Brucher, Perrot and Duchesnay}]{pedregosa2011}
\bibinfo{author}{Pedregosa, F.}, \bibinfo{author}{Varoquaux, G.},
  \bibinfo{author}{Gramfort, A.}, \bibinfo{author}{Michel, V.},
  \bibinfo{author}{Thirion, B.}, \bibinfo{author}{Grisel, O.},
  \bibinfo{author}{Blondel, M.}, \bibinfo{author}{Prettenhofer, P.},
  \bibinfo{author}{Weiss, R.}, \bibinfo{author}{Dubourg, V.},
  \bibinfo{author}{Vanderplas, J.}, \bibinfo{author}{Passos, A.},
  \bibinfo{author}{Cournapeau, D.}, \bibinfo{author}{Brucher, M.},
  \bibinfo{author}{Perrot, M.}, \bibinfo{author}{Duchesnay, E.},
  \bibinfo{year}{2011}.
\newblock \bibinfo{title}{Scikit-learn: Machine learning in {P}ython}.
\newblock \bibinfo{journal}{Journal of Machine Learning Research}
  \bibinfo{volume}{12}, \bibinfo{pages}{2825--2830}.
%Type = Misc
\bibitem[{{Polsterer} et~al.(2019){Polsterer}, {Gieseke} and
  {Doser}}]{polsterer2019}
\bibinfo{author}{{Polsterer}, K.L.}, \bibinfo{author}{{Gieseke}, F.},
  \bibinfo{author}{{Doser}, B.}, \bibinfo{year}{2019}.
\newblock \bibinfo{title}{{PINK: Parallelized rotation and flipping INvariant
  Kohonen maps}}.
\newblock \href{http://arxiv.org/abs/1910.001}{\tt arXiv:1910.001}.
%Type = Misc
\bibitem[{Reis et~al.(2019)Reis, Rotman, Poznanski, Prochaska and
  Wolf}]{reis2019}
\bibinfo{author}{Reis, I.}, \bibinfo{author}{Rotman, M.},
  \bibinfo{author}{Poznanski, D.}, \bibinfo{author}{Prochaska, J.X.},
  \bibinfo{author}{Wolf, L.}, \bibinfo{year}{2019}.
\newblock \bibinfo{title}{Effectively using unsupervised machine learning in
  next generation astronomical surveys}.
\newblock \href{http://arxiv.org/abs/1911.06823}{\tt arXiv:1911.06823}.
%Type = Article
\bibitem[{Richards et~al.(2011)Richards, Starr, Brink, Miller, Bloom, Butler,
  Berian~James, Long and Rice}]{Richards_2011}
\bibinfo{author}{Richards, J.W.}, \bibinfo{author}{Starr, D.L.},
  \bibinfo{author}{Brink, H.}, \bibinfo{author}{Miller, A.A.},
  \bibinfo{author}{Bloom, J.S.}, \bibinfo{author}{Butler, N.R.},
  \bibinfo{author}{Berian~James, J.}, \bibinfo{author}{Long, J.P.},
  \bibinfo{author}{Rice, J.}, \bibinfo{year}{2011}.
\newblock \bibinfo{title}{Active learning to overcome sample selection bias:
  Application to photometric variable star classification}.
\newblock \bibinfo{journal}{The Astrophysical Journal} \bibinfo{volume}{744},
  \bibinfo{pages}{192}.
\newblock \URLprefix \url{http://dx.doi.org/10.1088/0004-637X/744/2/192},
  \DOIprefix\doi{10.1088/0004-637x/744/2/192}.
%Type = Article
\bibitem[{{Roberts} et~al.(2019){Roberts}, {Bassett} and
  {Lochner}}]{roberts2019}
\bibinfo{author}{{Roberts}, E.}, \bibinfo{author}{{Bassett}, B.A.},
  \bibinfo{author}{{Lochner}, M.}, \bibinfo{year}{2019}.
\newblock \bibinfo{title}{{Bayesian Anomaly Detection and Classification}}.
\newblock \bibinfo{journal}{International Journal for Hybrid Intelligent
  Systems (in press)} ,
  \bibinfo{pages}{arXiv:1902.08627}\href{http://arxiv.org/abs/1902.08627}{\tt
  arXiv:1902.08627}.
%Type = Techreport
\bibitem[{Settles(2009)}]{settles2009}
\bibinfo{author}{Settles, B.}, \bibinfo{year}{2009}.
\newblock \bibinfo{title}{Active Learning Literature Survey}.
\newblock \bibinfo{type}{Computer Sciences Technical Report}
  \bibinfo{number}{1648}. University of Wisconsin--Madison.
\newblock \URLprefix
  \url{http://axon.cs.byu.edu/~martinez/classes/778/Papers/settles.activelearning.pdf}.
%Type = Article
\bibitem[{Solarz et~al.(2017)Solarz, Bilicki, Gromadzki, Pollo, Durkalec and
  Wypych}]{Solarz2017}
\bibinfo{author}{Solarz, A.}, \bibinfo{author}{Bilicki, M.},
  \bibinfo{author}{Gromadzki, M.}, \bibinfo{author}{Pollo, A.},
  \bibinfo{author}{Durkalec, A.}, \bibinfo{author}{Wypych, M.},
  \bibinfo{year}{2017}.
\newblock \bibinfo{title}{Automated novelty detection in the wise survey with
  one-class support vector machines}.
\newblock \bibinfo{journal}{Astronomy \& Astrophysics} \bibinfo{volume}{606},
  \bibinfo{pages}{A39}.
\newblock \URLprefix \url{http://dx.doi.org/10.1051/0004-6361/201730968},
  \DOIprefix\doi{10.1051/0004-6361/201730968}.
%Type = Inproceedings
\bibitem[{Spackman(1989)}]{spackman1989}
\bibinfo{author}{Spackman, K.A.}, \bibinfo{year}{1989}.
\newblock \bibinfo{title}{Signal detection theory: Valuable tools for
  evaluating inductive learning}, in: \bibinfo{booktitle}{Proceedings of the
  sixth international workshop on Machine learning},
  \bibinfo{organization}{Morgan Kaufmann Publishers Inc.}. pp.
  \bibinfo{pages}{160--163}.
%Type = Article
\bibitem[{Vafaei~Sadr et~al.(2019)Vafaei~Sadr, Bassett and Kunz}]{sadr2019}
\bibinfo{author}{Vafaei~Sadr, A.}, \bibinfo{author}{Bassett, B.A.},
  \bibinfo{author}{Kunz, M.}, \bibinfo{year}{2019}.
\newblock \bibinfo{title}{A flexible framework for anomaly detection via
  dimensionality reduction}.
\newblock \bibinfo{journal}{2019 6th International Conference on Soft Computing
  \& Machine Intelligence (ISCMI)} \URLprefix
  \url{http://dx.doi.org/10.1109/ISCMI47871.2019.9004400},
  \DOIprefix\doi{10.1109/iscmi47871.2019.9004400},
  \href{http://arxiv.org/abs/1909.04060}{\tt arXiv:1909.04060}.
%Type = Article
\bibitem[{Virtanen et~al.(2020)Virtanen, Gommers, Oliphant, Haberland, Reddy,
  Cournapeau, Burovski, Peterson, Weckesser, Bright, {van der Walt}, Brett,
  Wilson, Millman, Mayorov, Nelson, Jones, Kern, Larson, Carey, Polat, Feng,
  Moore, {VanderPlas}, Laxalde, Perktold, Cimrman, Henriksen, Quintero, Harris,
  Archibald, Ribeiro, Pedregosa, {van Mulbregt} and {SciPy 1.0
  Contributors}}]{virtanen2020}
\bibinfo{author}{Virtanen, P.}, \bibinfo{author}{Gommers, R.},
  \bibinfo{author}{Oliphant, T.E.}, \bibinfo{author}{Haberland, M.},
  \bibinfo{author}{Reddy, T.}, \bibinfo{author}{Cournapeau, D.},
  \bibinfo{author}{Burovski, E.}, \bibinfo{author}{Peterson, P.},
  \bibinfo{author}{Weckesser, W.}, \bibinfo{author}{Bright, J.},
  \bibinfo{author}{{van der Walt}, S.J.}, \bibinfo{author}{Brett, M.},
  \bibinfo{author}{Wilson, J.}, \bibinfo{author}{Millman, K.J.},
  \bibinfo{author}{Mayorov, N.}, \bibinfo{author}{Nelson, A.R.J.},
  \bibinfo{author}{Jones, E.}, \bibinfo{author}{Kern, R.},
  \bibinfo{author}{Larson, E.}, \bibinfo{author}{Carey, C.J.},
  \bibinfo{author}{Polat, {\.I}.}, \bibinfo{author}{Feng, Y.},
  \bibinfo{author}{Moore, E.W.}, \bibinfo{author}{{VanderPlas}, J.},
  \bibinfo{author}{Laxalde, D.}, \bibinfo{author}{Perktold, J.},
  \bibinfo{author}{Cimrman, R.}, \bibinfo{author}{Henriksen, I.},
  \bibinfo{author}{Quintero, E.A.}, \bibinfo{author}{Harris, C.R.},
  \bibinfo{author}{Archibald, A.M.}, \bibinfo{author}{Ribeiro, A.H.},
  \bibinfo{author}{Pedregosa, F.}, \bibinfo{author}{{van Mulbregt}, P.},
  \bibinfo{author}{{SciPy 1.0 Contributors}}, \bibinfo{year}{2020}.
\newblock \bibinfo{title}{{{SciPy} 1.0: Fundamental Algorithms for Scientific
  Computing in Python}}.
\newblock \bibinfo{journal}{Nature Methods} \bibinfo{volume}{17},
  \bibinfo{pages}{261--272}.
\newblock \DOIprefix\doi{10.1038/s41592-019-0686-2}.
%Type = Article
\bibitem[{Walmsley et~al.(2019)Walmsley, Smith, Lintott, Gal, Bamford,
  Dickinson, Fortson, Kruk, Masters, Scarlata and et~al.}]{Walmsley_2019}
\bibinfo{author}{Walmsley, M.}, \bibinfo{author}{Smith, L.},
  \bibinfo{author}{Lintott, C.}, \bibinfo{author}{Gal, Y.},
  \bibinfo{author}{Bamford, S.}, \bibinfo{author}{Dickinson, H.},
  \bibinfo{author}{Fortson, L.}, \bibinfo{author}{Kruk, S.},
  \bibinfo{author}{Masters, K.}, \bibinfo{author}{Scarlata, C.},
  \bibinfo{author}{et~al.}, \bibinfo{year}{2019}.
\newblock \bibinfo{title}{Galaxy zoo: probabilistic morphology through bayesian
  cnns and active learning}.
\newblock \bibinfo{journal}{Monthly Notices of the Royal Astronomical Society}
  \bibinfo{volume}{491}, \bibinfo{pages}{1554–1574}.
\newblock \URLprefix \url{http://dx.doi.org/10.1093/mnras/stz2816},
  \DOIprefix\doi{10.1093/mnras/stz2816}.
%Type = Article
\bibitem[{Webb et~al.(2020)Webb, Lochner, Muthukrishna, Cooke, Flynn, Mahabal,
  Goode, Andreoni, Pritchard and Abbott}]{webb2020}
\bibinfo{author}{Webb, S.}, \bibinfo{author}{Lochner, M.},
  \bibinfo{author}{Muthukrishna, D.}, \bibinfo{author}{Cooke, J.},
  \bibinfo{author}{Flynn, C.}, \bibinfo{author}{Mahabal, A.},
  \bibinfo{author}{Goode, S.}, \bibinfo{author}{Andreoni, I.},
  \bibinfo{author}{Pritchard, T.}, \bibinfo{author}{Abbott, T.M.C.},
  \bibinfo{year}{2020}.
\newblock \bibinfo{title}{Unsupervised machine learning for transient discovery
  in deeper, wider, faster light curves}.
\newblock \bibinfo{journal}{Monthly Notices of the Royal Astronomical Society}
  \bibinfo{volume}{498}, \bibinfo{pages}{3077–3094}.
\newblock \URLprefix \url{http://dx.doi.org/10.1093/mnras/staa2395},
  \DOIprefix\doi{10.1093/mnras/staa2395}.
%Type = Article
\bibitem[{{Willett} et~al.(2013){Willett}, {Lintott}, {Bamford}, {Masters},
  {Simmons}, {Casteels}, {Edmondson}, {Fortson}, {Kaviraj}, {Keel}, {Melvin},
  {Nichol}, {Raddick}, {Schawinski}, {Simpson}, {Skibba}, {Smith} and
  {Thomas}}]{willett2013}
\bibinfo{author}{{Willett}, K.W.}, \bibinfo{author}{{Lintott}, C.J.},
  \bibinfo{author}{{Bamford}, S.P.}, \bibinfo{author}{{Masters}, K.L.},
  \bibinfo{author}{{Simmons}, B.D.}, \bibinfo{author}{{Casteels}, K.R.V.},
  \bibinfo{author}{{Edmondson}, E.M.}, \bibinfo{author}{{Fortson}, L.F.},
  \bibinfo{author}{{Kaviraj}, S.}, \bibinfo{author}{{Keel}, W.C.},
  \bibinfo{author}{{Melvin}, T.}, \bibinfo{author}{{Nichol}, R.C.},
  \bibinfo{author}{{Raddick}, M.J.}, \bibinfo{author}{{Schawinski}, K.},
  \bibinfo{author}{{Simpson}, R.J.}, \bibinfo{author}{{Skibba}, R.A.},
  \bibinfo{author}{{Smith}, A.M.}, \bibinfo{author}{{Thomas}, D.},
  \bibinfo{year}{2013}.
\newblock \bibinfo{title}{{Galaxy Zoo 2: detailed morphological classifications
  for 304 122 galaxies from the Sloan Digital Sky Survey}}.
\newblock \bibinfo{journal}{\mnras} \bibinfo{volume}{435},
  \bibinfo{pages}{2835--2860}.
\newblock \DOIprefix\doi{10.1093/mnras/stt1458},
  \href{http://arxiv.org/abs/1308.3496}{\tt arXiv:1308.3496}.
%Type = Misc
\bibitem[{Willett et~al.(2013)Willett, Lintott, Bamford, Masters, Simmons,
  Casteels, Edmonson, Fortson, Kaviraj, Keel, Melvin, Nichol, Raddick,
  Schawinski, Simpson, Skibba, Smith and Thomas}]{willett2013zenodo}
\bibinfo{author}{Willett, K.W.}, \bibinfo{author}{Lintott, C.J.},
  \bibinfo{author}{Bamford, S.P.}, \bibinfo{author}{Masters, K.L.},
  \bibinfo{author}{Simmons, B.D.}, \bibinfo{author}{Casteels, K.R.V.},
  \bibinfo{author}{Edmonson, E.M.}, \bibinfo{author}{Fortson, L.F.},
  \bibinfo{author}{Kaviraj, S.}, \bibinfo{author}{Keel, W.C.},
  \bibinfo{author}{Melvin, T.}, \bibinfo{author}{Nichol, R.C.},
  \bibinfo{author}{Raddick, M.J.}, \bibinfo{author}{Schawinski, K.},
  \bibinfo{author}{Simpson, R.J.}, \bibinfo{author}{Skibba, R.A.},
  \bibinfo{author}{Smith, A.M.}, \bibinfo{author}{Thomas, D.},
  \bibinfo{year}{2013}.
\newblock \bibinfo{title}{Galaxy zoo 2: Images from original sample}.
\newblock \URLprefix \url{https://doi.org/10.5281/zenodo.3565489},
  \DOIprefix\doi{10.5281/zenodo.3565489}.
%Type = Article
\bibitem[{Wolpert(1996)}]{Wolpert1996}
\bibinfo{author}{Wolpert, D.H.}, \bibinfo{year}{1996}.
\newblock \bibinfo{title}{The lack of a priori distinctions between learning
  algorithms}.
\newblock \bibinfo{journal}{Neural Computation} , \bibinfo{pages}{1341--1390}.
%Type = Article
\bibitem[{Wolpert and Macready(1997)}]{Wolpert1997}
\bibinfo{author}{Wolpert, D.H.}, \bibinfo{author}{Macready, W.G.},
  \bibinfo{year}{1997}.
\newblock \bibinfo{title}{No free lunch theorems for optimization}.
\newblock \bibinfo{journal}{IEEE transactions on evolutionary computation}
  \bibinfo{volume}{1}, \bibinfo{pages}{67--82}.
%Type = Misc
\bibitem[{Škoda et~al.(2020)Škoda, Podsztavek and Tvrdík}]{skoda2020}
\bibinfo{author}{Škoda, P.}, \bibinfo{author}{Podsztavek, O.},
  \bibinfo{author}{Tvrdík, P.}, \bibinfo{year}{2020}.
\newblock \bibinfo{title}{Active deep learning method for the discovery of
  objects of interest in large spectroscopic surveys}.
\newblock \href{http://arxiv.org/abs/2009.03219}{\tt arXiv:2009.03219}.

\end{thebibliography}
\newpage
\appendix
\section{\astronomaly visual description}
\label{sec:astronomaly_flowchart}

\ml{Figure \ref{fig:flowchart} shows a visual representation of the \astronomaly framework. The majority of the backend operates as a simple linear pipeline typical of machine learning applications. Data management allows data to be read into the particular format \astronomaly requires. Data may be read in at once, if it is a small dataset, or on demand as is the case for Galaxy Zoo data which is too large to fit into memory. Data can be preprocessed (which could include resizing, contrast enhancement etc.) before feature extraction. If the feature space is large, \astronomaly includes some dimensionality reduction routines such as principal component analysis. Finally, as many machine learning algorithms require features to be rescaled to zero mean and unit variance, a postprocessing step is generally implemented.}

\noindent
\begin{minipage}{1\linewidth}
    \centering
    \gap
    \includegraphics[width=0.99\columnwidth]{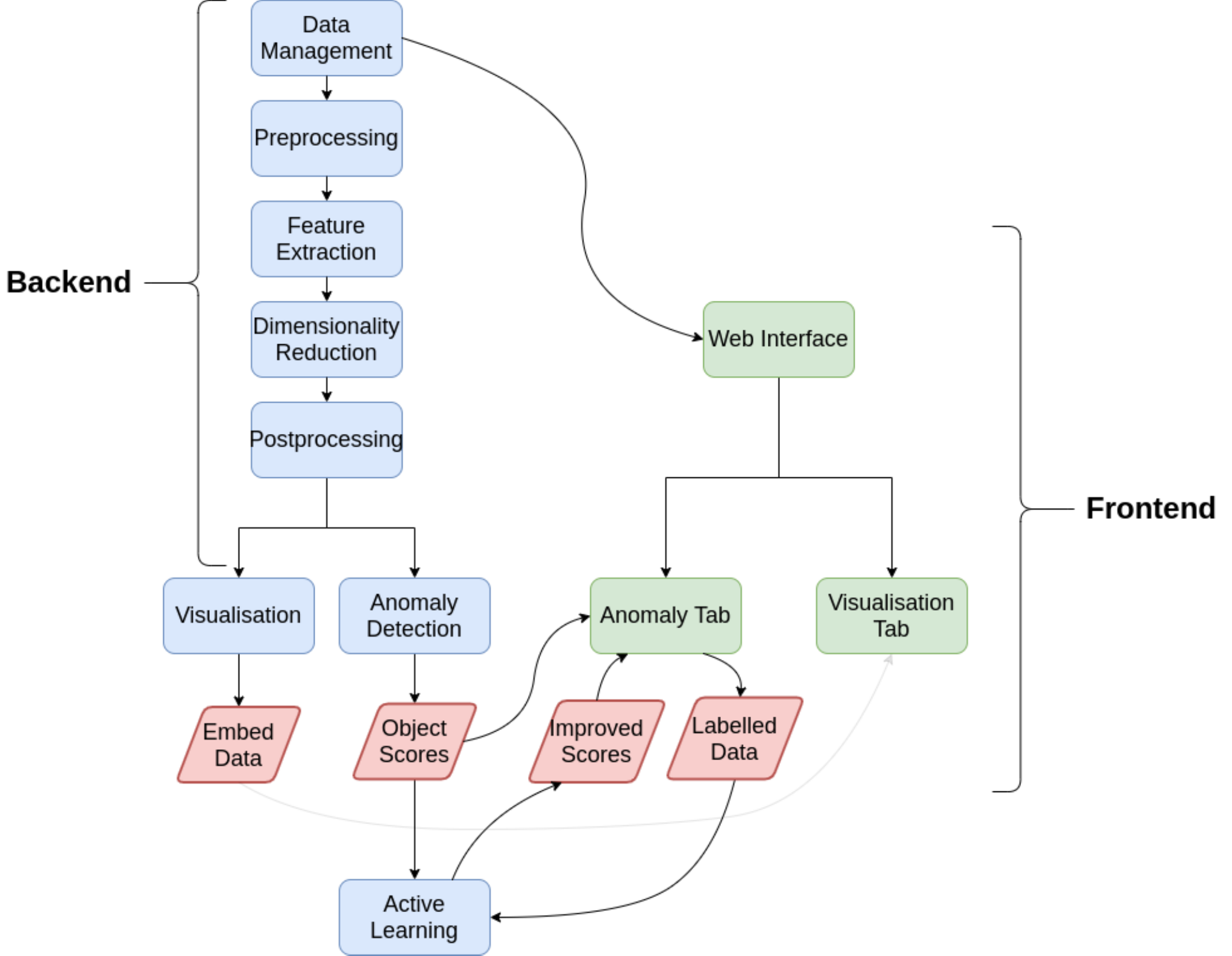}
    \captionof{figure}{Diagram illustrating the structure of the \astronomaly code. Each rectangle represents a component while a parallelogram is some output passed between the backend and frontend. Further details can be found in the \astronomaly documentation.}
    \label{fig:flowchart}
\end{minipage}

\ml{After the linear pipeline, the user may interact with the data in two ways. The first is visualisation, for which t-SNE was used in this paper, which allows the user to analyse the feature space. The second, which is the main component of \astronomaly, is anomaly detection. An anomaly detection algorithm is run on all objects and they are then ordered from most to least anomalous and presented visually to the user in the frontend. The user can then interactively iterate through the dataset and examine the most anomalous objects. By making use of the user-provided labels, \astronomaly can then apply active learning to recommend more interesting objects to the user.}

\ml{The interactive active learning process proceeds as follows:}
\ml{\begin{itemize}
    \item Run anomaly detection on all data
\item Rank objects from most to least anomalous, present to user in frontend
\item The user labels as many objects as they wish on a scale of 0-5 based on how interesting the anomaly appears
\item The user then presses the “retrain” button to initiate active learning
\item \astronomaly runs a regression algorithm to predict the user relevance score for all objects in the dataset, based on the limited training set 
\item The active learning score (Equation \ref{eq:active_learning}) is computed for every object in the dataset, which incorporates the machine learning anomaly score, the user relevance and an uncertainty term
\item The list of anomalies presented to the user is now ranked by the updated active learning score
\item The user may then repeat the process to improve the active learning by labelling more data
\end{itemize}}

\section{Details of the ellipse-fitting feature extraction technique}
\label{sec:ellipse}
For this work, we developed a new feature extraction technique that is sensitive to unusual galaxy morphology, reasonably robust to noise, rotation invariant and theoretically independent of the size of the object in question (in practice we were limited by the fixed size of the cutouts in Section \ref{sec:galaxy_zoo}).

Figure \ref{fig:galaxy_ellipse} illustrates the feature extraction process, which is as follows:
\begin{enumerate}
    \item We first apply sigma clipping to the image using Astropy \citep{astropy2013, astropy2018}. Sigma clipping is an iterative procedure that removes the pixels above a threshold value (such as $3\sigma$) from the image, determines the standard deviation of the remaining pixels and repeats the process until the standard deviation has stabilised, which means the remaining pixels are noise and the determined standard deviation is the ``true'' one.
    \item We use the sigma-clipped standard deviation, apply a $3\sigma$ threshold and use OpenCV \citep{bradski2000} to extract the contour corresponding to this threshold that encompasses the centre of the image (where the source is assumed to be). This contour is used to cut out the central source, thus removing noise and bright nearby sources. The second row of Figure \ref{fig:galaxy_ellipse} shows examples of images after sigma-clipping has been applied.
    \item We determine the threshold brightness values corresponding to the following percentiles: 90, 80, 70, 60, 50, 0. OpenCV is used to extract the contours corresponding to each of these thresholds, as can be seen in the third row of Figure \ref{fig:galaxy_ellipse}.
    \item An ellipse is fitted to each contour (using OpenCV) and the parameters are stored. The fourth row of Figure \ref{fig:galaxy_ellipse} shows the fitted ellipses.
    \item Before the ellipse parameters can be used as features, they must be transformed to ensure they are rotation invariant. We use the following parameters for each ellipse:
\begin{itemize}
    \item {\bf Residual} - the sum of the difference between the fitted ellipse and the contours.
    \item {\bf Offset} - the Euclidean distance between the centre of the ellipse and the centre of the $90^{\rm th}$ percentile ellipse (which is usually the innermost ellipse).
    \item {\bf Aspect} - the ratio of the major to the minor axis of the ellipse, divided by the aspect ratio of the $90^{\rm th}$ percentile ellipse.
    \item {\bf Theta} - the absolute value of the rotation angle of the ellipse minus that of the $90^{\rm th}$ percentile ellipse, in degrees.
\end{itemize}
    \item This results in 21 parameters (only the residual is used as a feature for the $90^{\rm th}$ percentile ellipse, since all other parameters are relative to this ellipse). By inspecting the third row of Figure \ref{fig:galaxy_ellipse}, it is clear that for normal galaxies the ellipses tend to line up and the parameters will be similar. For anomalous objects however, the ellipses will often lie at very different angles, be offset from each other and have different aspect ratios. The ellipses will also be a poor fit to the contours (residual).
    \item For a small percentage of data, this procedure can fail. Usually this is because of an object in the field so bright, the central source falls below the $3\sigma$ threshold and is not extracted in this phase of the analysis. We do not attempt anomaly detection for these objects (around 1\% of the data).
\end{enumerate}

\newpage
\noindent
\begin{minipage}{1\linewidth}
    \begin{center}
    \begin{minipage}{0.9\linewidth}
        \begin{minipage}{0.33\linewidth}
            \includegraphics[width=1\linewidth]{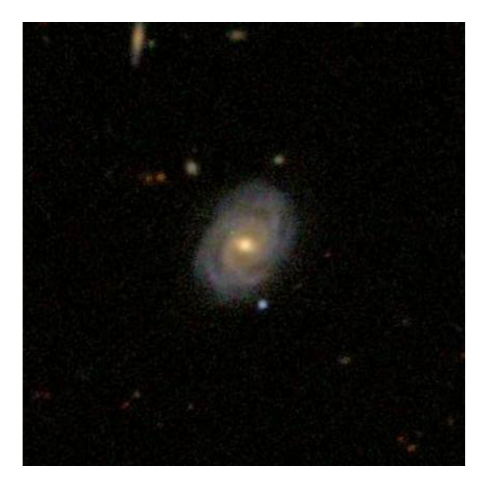}
        \end{minipage}
        \begin{minipage}{0.33\linewidth}
            \includegraphics[width=1\linewidth]{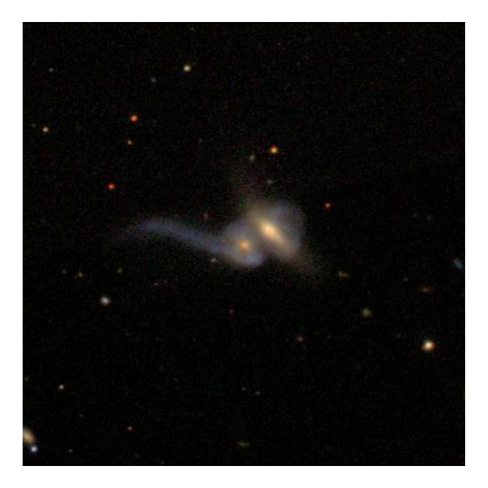}
        \end{minipage}
        \begin{minipage}{0.33\linewidth}
            \includegraphics[width=1\linewidth]{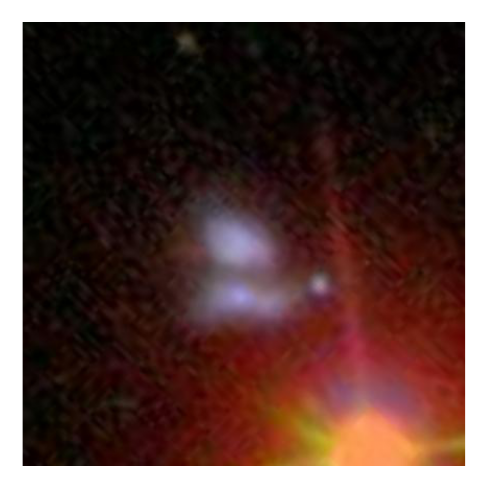}
        \end{minipage}
    \end{minipage}
    
    \begin{minipage}{0.9\linewidth}
        \begin{minipage}{0.33\linewidth}
            \includegraphics[width=1\linewidth]{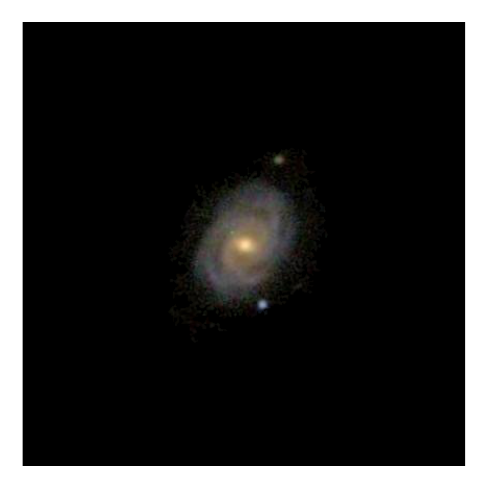}
        \end{minipage}
        \begin{minipage}{0.33\linewidth}
            \includegraphics[width=1\linewidth]{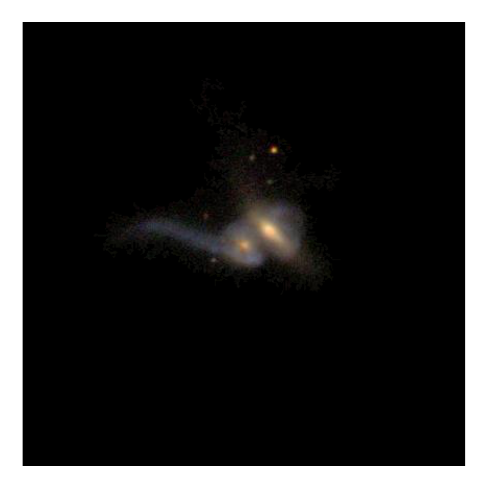}
        \end{minipage}
        \begin{minipage}{0.33\linewidth}
            \includegraphics[width=1\linewidth]{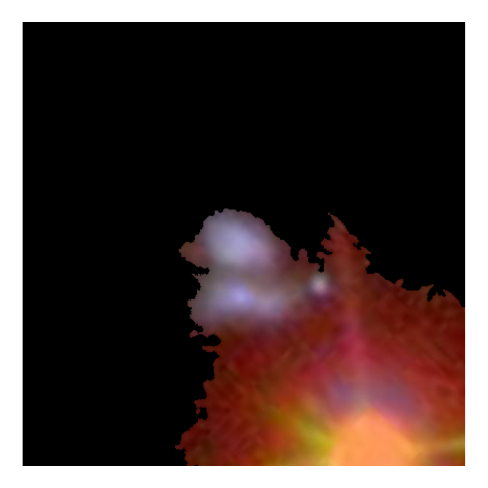}
        \end{minipage}
    \end{minipage}
    
    \begin{minipage}{0.9\linewidth}
        \begin{minipage}{0.33\linewidth}
            \includegraphics[width=1\linewidth]{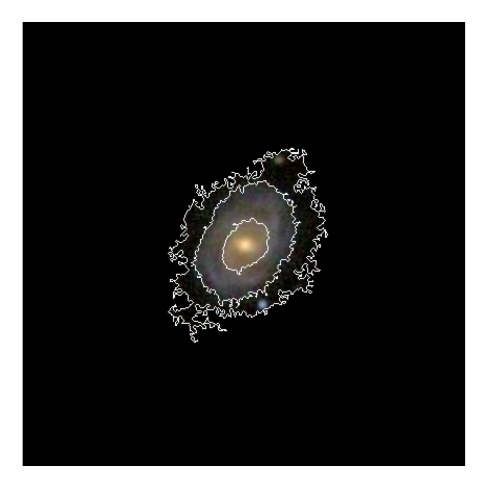}
        \end{minipage}
        \begin{minipage}{0.33\linewidth}
            \includegraphics[width=1\linewidth]{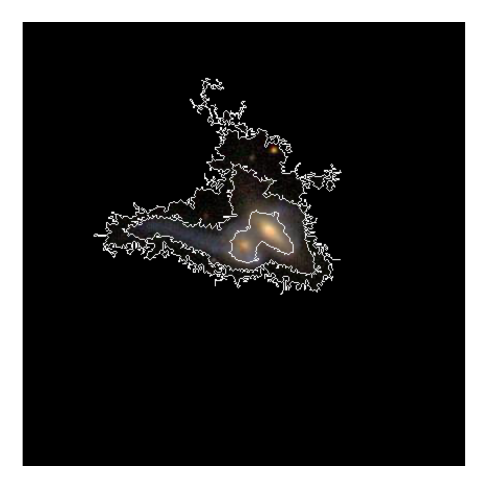}
        \end{minipage}
        \begin{minipage}{0.33\linewidth}
            \includegraphics[width=1\linewidth]{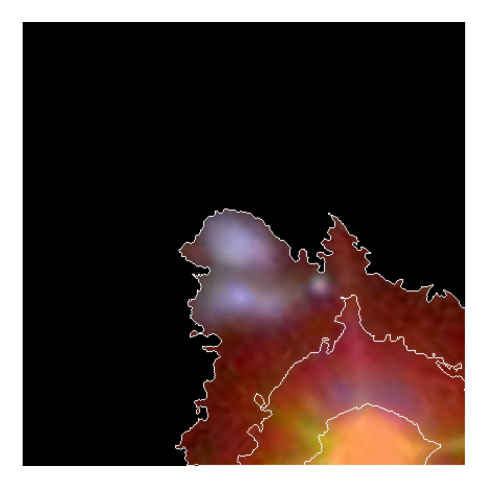}
        \end{minipage}
    \end{minipage}
    
    \begin{minipage}{0.9\linewidth}
        \begin{minipage}{0.33\linewidth}
            \includegraphics[width=1\linewidth]{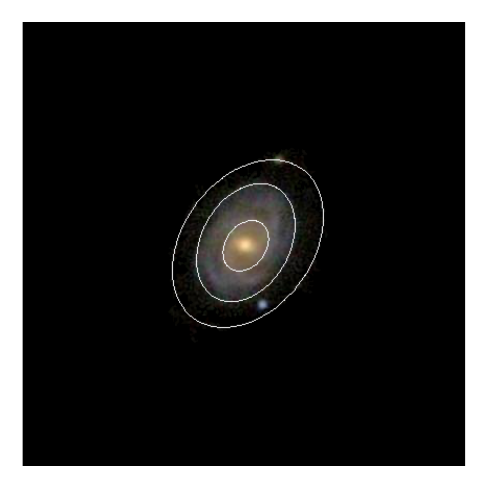}
            \captionof{subfigure}{Normal galaxy}
        \end{minipage}
        \begin{minipage}{0.33\linewidth}
            \includegraphics[width=1\linewidth]{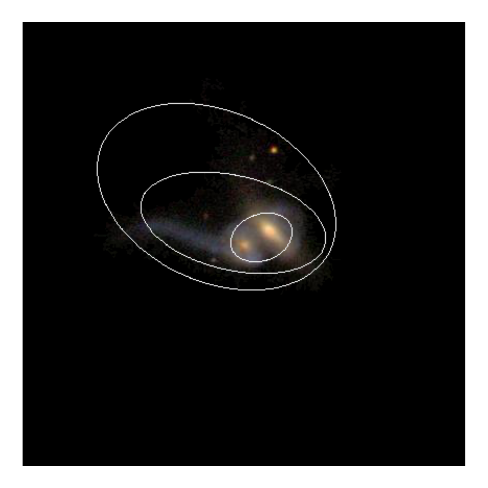}
            \captionof{subfigure}{Anomalous galaxy}
        \end{minipage}
        \begin{minipage}{0.33\linewidth}
            \includegraphics[width=1\linewidth]{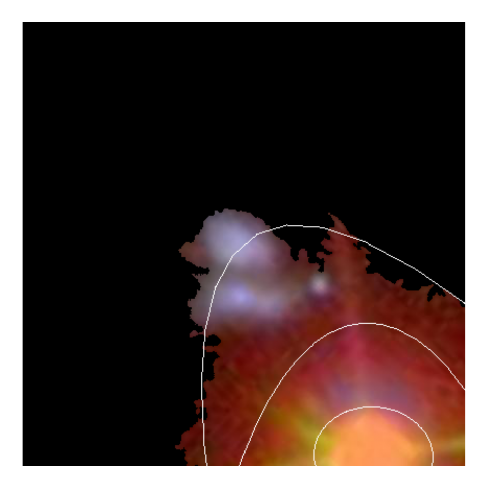}
            \captionof{subfigure}{Contaminating star}
        \end{minipage}
    \end{minipage}
    
    \end{center}
    \captionof{figure}{Illustration of the ellipse fitting feature extraction method. The top row shows the input image, the second shows the image after sigma clipping, the third row shows the contours extracted for 3 different thresholds and the fourth shows the corresponding fitted ellipses. The procedure is demonstrated for (a) a normal galaxy, (b) an anomalous object showing how the ellipses mis-align and thus will stand out as anomalous and (c) an image with a contaminating star that is not removed by sigma-clipping and hence artificially distorts the ellipses.}
    \label{fig:galaxy_ellipse}
\end{minipage}

\section{Computational performance}
\label{sec:computational}
Computational performance is critical for \astronomaly to be successfully run on large datasets. Here we report the timing analysis for the two datasets used in this paper: the Galazy Zoo data  (Table \ref{tab:galaxy_compute}) and the simulations (Table \ref{tab:sim_compute}). All analysis was done on a standard laptop (Intel i7 1.8GHz processor, 16GB RAM) using the processor only (no GPU acceleration) and without parallelisation. 

An important point to note is that by far the most computationally expensive task is feature extraction. Anomaly detection, active learning and visualisation are negligible in comparison. This is frequently the case for machine learning problems. The Galaxy Zoo data represents one of the most computationally demanding datasets that one could imagine running \astronomaly on. The images are large (400x400 pixels), likely much larger than would be used for most data, so represents the probable upper limit in compute time for this feature extraction method. With this in mind and looking at the timing in Table \ref{tab:galaxy_compute}, it is clear that \astronomaly can scale well to much larger datasets. Furthermore, while \astronomaly does not currently make use of parallelisation (although it is constructed with that future goal in mind), feature extraction methods can be trivially parallelised as data can be split up and processed between processors. 

A final point is to note that many problems will be memory limited, not processing power limited. A large dataset of astronomical images will, in general, not fit into memory. \astronomaly gets around this (when working with fits images as opposed to the cutouts in the Galaxy Zoo example here) by only extracting samples on demand, such as during feature extraction, and avoiding loading the entire dataset into memory. However there is still room for improvement in balancing memory usage and running time. \astronomaly as it stands should work well for the majority of datasets astronomers may wish to work with. 

\noindent
\begin{minipage}{1\linewidth}
    \centering
    \begin{tabular}{ccc}
    \hline
    Task & Notes on method & Time (seconds) \\
    \hline
    Feature extraction & Sigma-clipping and ellipse fitting &  56.01 min\\
    Anomaly detection & Isolation forest &  2.20\\
    Active learning & 200 labelled objects used & 3.85 \\
    Visualisation (t-SNE) & Subset of 2000 objects  & 40.24 \\
    \hline
    \end{tabular}
    \captionof{table}{Timing analysis for the Galaxy Zoo data described in Section \ref{sec:galaxy_zoo}, which is a dataset of 61578 400x400 pixel images. The feature extraction method results in 21 features.}
    \label{tab:galaxy_compute}
    \vspace{20pt}
\end{minipage}

\noindent
\begin{minipage}{1\linewidth}
    \centering
    \begin{tabular}{ccc}
    \hline
    Task & Notes on method & Time (seconds) \\
    \hline
    Feature extraction & Raw data used as features & N/A\\
    Anomaly detection & Local outlier factor & 199.91 \\
    Active learning & 100 labelled objects used & 4.31 \\
    Visualisation (t-SNE) & Subset of 2000 objects  & 40.61 \\
    \hline
    \end{tabular}
    \captionof{table}{Timing analysis for the simulations described in Section \ref{sec:simulations}, which is a dataset of 100 features and just under 50000 objects.}
    \label{tab:sim_compute}
    \vspace{20pt}
\end{minipage}

\end{document}